\documentclass[11pt,a4paper]{article}
\pdfoutput=1

\usepackage{jheppub}
\usepackage{amsmath}
\usepackage{amssymb}
\usepackage{graphics}
\usepackage[active]{srcltx}
\usepackage{pdfsync}
\usepackage{shuffle}
\usepackage{slashed}
\usepackage{hyperref}
\usepackage{subfigure}

\newcommand{\bea}{\begin{eqnarray}}
\newcommand{\eea}{\end{eqnarray}}
\newcommand{\nn}{\nonumber \\}

\def\W #1{\widetilde{#1}}

\def\eref#1{(\ref{#1})}

\def\a{{\alpha}}

\def\b{{\beta}}

\allowdisplaybreaks

\title{Towards tree Yang-Mills and Yang-Mills-scalar amplitudes with higher-derivative interactions}
 \author[a]{Kang Zhou} \author[b,c]{Chang Hu}

\affiliation[a]{Center for Gravitation and Cosmology, College of Physical Science and Technology, Yangzhou University,\\
No.180, Siwangting Road, Yangzhou, 225009, P.R. China}

\affiliation[b]{School of Fundamental Physics and Mathematical Sciences, Hangzhou Institute for Advanced Study, UCAS, Hangzhou 310024, China}

\affiliation[c]{University of Chinese Academy of Sciences, Beijing 100049, China}

\emailAdd{zhoukang@yzu.edu.cn} \emailAdd{isiahalbert@126.com}

\date{\today}
\abstract{In our recent works, a new approach for constructing tree amplitudes, based on exploiting soft behaviors,
was proposed. In this paper, we extend this approach to effective theories for gluons which incorporate higher-derivative
interactions. By applying our method, we construct tree Yang-Mills (YM) and Yang-Mills-scalar (YMS) amplitudes with the single insertion
of $F^3$ local operator, as well as the YM amplitudes those receive contributions from both $F^3$ and $F^4$ operators.
All results are represented as universal expansions to appropriate basis. We also conjecture a compact general formula for tree YM amplitudes with
higher mass dimension, which allows us to generate them from ordinary YM amplitudes, and discuss the consistent factorizations of the
conjectured formula.
}


\begin{document}

\maketitle \flushbottom

\section{Introduction}
\label{sec-intro}

The modern S-matrix program aims to construct scattering amplitudes directly, solely relying on physical principles like Lorentz invariance and unitarity, without traditional Lagrangians or equations of motion (see for reviews in \cite{Elvang:2013cua,Cheung:2017pzi}). Notably, the Britto-Cachazo-Feng-Witten (BCFW) recursion relation \cite{Britto:2004ap,Britto:2005fq} is a prime example, utilizing on-shell information to recursively build higher-point amplitudes from lower-point ones. This paper focuses on novel approaches for computing tree amplitudes, particularly those centered around the inversion of soft theorems.

Soft theorems show the universal behavior of scattering amplitudes in the "soft limit," where the momentum of an external particle approaches zero, denoted by $\tau\to 0$ with a rescaling of massless external momentum $k_i\to\tau k_i$. Initially proposed for photons and gravitons \cite{Low:1958sn,Weinberg:1965nx}, they garnered renewed attention since 2014 \cite{Cachazo:2014fwa,Casali:2014xpa,Schwab:2014xua,Afkhami-Jeddi:2014fia}, particularly due to their extension to higher orders in gravity (GR) and Yang-Mills (YM) theory. Subsequently, the study of soft theorems was broadened to encompass string theory and loop levels \cite{Bern:2014oka,He:2014bga,Cachazo:2014dia,Bianchi:2014gla,Sen:2017nim}. Soft theorems have been utilized to construct tree amplitudes independently of Lagrangians, including through programs such as the inverse soft limit (ISL) for YM, GR, and Einstein-Yang-Mills (EYM) amplitudes, and the construction of amplitudes for various scalar effective theories based on the soft behavior known as Adler zero \cite{Nguyen:2009jk,Boucher-Veronneau:2011rwd,Rodina:2018pcb,Ma:2022qja,Cheung:2014dqa,Cheung:2015ota,Luo:2015tat,Elvang:2018dco}.

The ISL program, rooted in BCFW recursion, efficiently generates $n$-point amplitudes from $(n-1)$-point ones for $4$-dimensional tree amplitudes in spinor-helicity representation. However, it lacks insights into general amplitudes with arbitrary numbers of external legs. In recent works, we proposed a new bottom-up approach for inverting soft theorems \cite{Zhou:2022orv,Wei:2023yfy,Hu:2023lso,Du:2024dwm}. Our technique is insensitive to specific deformations of external momenta and the dimension of space-time, yielding general tree amplitudes represented through universal expansions in appropriate bases. Effective for single- and multiple-trace Yang-Mills-scalar (YMS), YM, EYM, and GR amplitudes, this method offers significant advantages\cite{Zhou:2022orv,Wei:2023yfy,Hu:2023lso,Du:2024dwm}. Taking tree YM amplitudes as an example, our new approach begins by bootstrapping the $3$-point YM amplitudes, expressing them as combinations of $3$-point bi-adjoint-scalar (BAS) amplitudes. Subsequently, the $4$-point YM amplitudes are constructed from $3$-point ones by inverting the subleading soft theorem for gluons, resulting in expansions to $4$-point BAS amplitudes. This construction reveals a recursive pattern that straightforwardly leads to the expansion formula of general YM amplitudes with arbitrary numbers of external gluons, aligning with expansions found in the literature via alternative methods \cite{Fu:2017uzt,Feng:2019tvb}. Theoretically, this expansion elucidates the connection between YM and BAS amplitudes at the tree level. Practically, they simplified the computation of YM amplitudes, since BAS amplitudes are much simpler than YM ones.

It is natural to inquire whether this new approach can be extended to a broader scope. In this paper, we leverage this method to construct tree YM and YMS amplitudes featuring higher-dimensional interactions. Higher-dimensional operators play crucial roles across various domains. They can emerge in the effective action of open strings and act as potential counter-terms for UV divergences at loop levels. Additionally, they are of considerable phenomenological interest as potential corrections to YM theory.

We start by constructing tree-level YM amplitudes with mass dimension ${\cal D}+2$, where ${\cal D}$ denotes the mass dimension of standard YM amplitudes.\footnote{It's worth noting that all amplitudes considered in this paper consist solely of the kinematic parts obtained by stripping away coupling constants and color orders. Therefore, the mass dimensions are calculated for these pure kinematic amplitudes.} Our approach begins with bootstrapping the unique $3$-point amplitudes, subsequently extending to higher-point ones by inverting the subleading soft theorem for gluons. While one might anticipate that the introduction of new higher-derivative interactions would alter the soft behavior of gluons in standard YM amplitudes, we argue and validate that the contribution from the new vertex vanishes at the subleading order under consideration. This implies that the subleading soft theorem for gluons in YM amplitudes persists, albeit with the inclusion of undetectable terms at this order. Eventually, the complete amplitudes are attained by considering symmetry, expressed through an universal expansion to ordinary YMS amplitudes, ultimately leading to an expansion in the BAS basis.

Although the construction described above is bottom-up and doesn't rely on information from a Lagrangian, the resulting amplitudes with mass dimension ${\cal D}+2$ turn out to be gluon amplitudes with a single insertion of the $F^3$ local operator. This operator serves as the leading correction to YM theory, arising from the $\alpha'$ expansion of bosonic string theory \cite{Polchinski:1998rq}. Additionally, it's considered a potential departure of gluon interactions from those in QCD, originating from new physics \cite{Simmons:1989zs,Simmons:1990dh,Cho:1993eu}.

The similar technic can also be applied to construct tree YMS amplitudes with mass dimension ${\cal D}'+2$, with ${\cal D}'$ the mass dimension of standard YMS amplitudes. These amplitudes depict gluons coupling to BAS scalars with a single insertion of the $F^3$ operator. We start by generating the lowest $4$-point amplitudes of this YMS theory by performing dimensional reduction on YM amplitudes with mass dimension ${\cal D}+2$. Then, we construct higher-point amplitudes by inverting soft theorems for BAS scalars and gluons. The resulting amplitudes are also expressed as an universal expansion in terms of ordinary YMS amplitudes.


Our method can also be applied to construct tree-level YM amplitudes with mass dimension ${\cal D}+4$ strictly.
The resulting amplitudes, are expressed as an universal expansion in terms of YMS amplitudes with mass dimension ${\cal D}'+2$. These amplitudes arise from two sectors of Feynman diagrams: one involves double insertions of $F^3$ operators, while the other includes single insertion of $F^4$ operators. Remarkably, our method seamlessly combines contributions from both sectors. The expansion formulas for YM amplitudes with mass dimensions ${\cal D}+2$ and ${\cal D}+4$ exhibit strong similarity, suggesting a general recursive pattern for generating amplitudes with mass dimension ${\cal D}+2h$ from those with mass dimension ${\cal D}+2(h-1)$. This observation motivates us to propose a general formula for YM amplitudes with arbitrary higher values of $h$. The conjectured formula enables the generation of YM amplitudes with higher mass dimensions from ordinary YM amplitudes. We recursively verify the consistent factorizations of the conjectured formula.

In the expansion formulas of YM and YMS amplitudes with insertions of higher-dimensional operators, each coefficient depends on only one of two orderings carried by the corresponding amplitude in the basis. As a consequence, the double copy structure and Bern-Carrasco-Johansson (BCJ) relations \cite{Kawai:1985xq,Bern:2008qj,Chiodaroli:2014xia,Johansson:2015oia,Johansson:2019dnu} are automatically satisfied. Furthermore, although the gauge invariance is not assumed as the priori input, our results manifest the gauge invariance for the polarization carried by any external gluon.

The reminder of this paper is organized as follows. For readers convenience, a rapid review for necessary background will be given in  section \ref{sec-preparations}. In section \ref{sec-YM+2}, we construct YM amplitudes with mass dimension ${\cal D}+2$, via our bottom-up method.  Then, in section \ref{sec-YMS+2}, we continue to construct YMS amplitudes with mass dimension ${\cal D}'+2$, along the similar line.
Section \ref{sec-YM+2h} aims to construct YM amplitudes with mass dimension ${\cal D}+4$, and give a conjecture for the general case with mass
dimension ${\cal D}+2h$.
We finally end with a brief summary in section \ref{sec-summary}.

\section{Brief review for soft behaviors of BAS, YMS and YM amplitudes}
\label{sec-preparations}

In this section, we review the soft behavior of scalars in tree BAS and YMS amplitudes, as well as the soft behavior of gluons in tree YMS and YM
amplitudes. In section \ref{subsec-introBASYMS}, we give a brief introduction for BAS and YMS amplitudes under consideration,
as well as various notations and conventions which will be frequently used throughout this paper. Then, in section \ref{subsec-introsoft},
we review the corresponding soft behaviors which play the crucial role in subsequent sections, and discuss their universality.

\subsection{BAS and YMS amplitudes}
\label{subsec-introBASYMS}

\begin{figure}
  \centering
  \includegraphics[width=6cm]{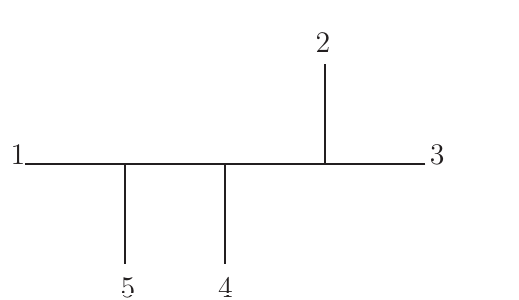}
   \includegraphics[width=6cm]{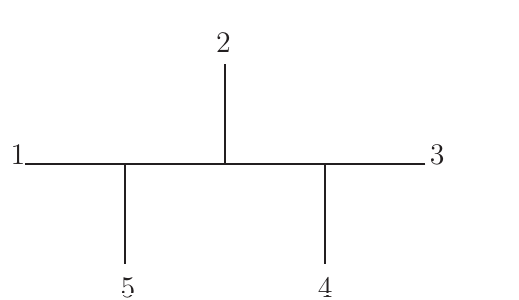}  \\
  \caption{Two $5$-point diagrams}\label{5p}
\end{figure}

The bi-adjoint scalar (BAS) theory describes a scenario where each massless scalar field $\phi^{Aa}$ carries two color indices. The Lagrangian is as follows:
\bea
{\cal L}_{\rm BAS}={1\over2}\,\partial_\mu\phi^{Aa}\,\partial^{\mu}\phi^{Aa}+{\lambda\over3!}\,F^{ABC}f^{abc}\,
\phi^{Aa}\phi^{Bb}\phi^{Cc}\,,
\eea
where $F^{ABC}={\rm tr}([T^A,T^B]T^C)$ and $f^{abc}={\rm tr}([T^a,T^b]T^c)$ are structure constants of two Lie groups respectively.
The tree amplitudes in this theory consist solely of propagators for massless scalars. The standard decomposition of group factors yields
\bea
{\cal A}_n=\sum_{\sigma\in{\cal S}_n\setminus Z_n}\,\sum_{\sigma'\in{\cal S}'_n\setminus Z'_n}\,
{\rm tr}[T^{A_{\sigma_1}},\cdots T^{A_{\sigma_n}}]\,{\rm tr}[T^{a_{\sigma'_1}}\cdots T^{a_{\sigma'_n}}]\,
{\cal A}_{\rm BAS}(\sigma_1,\cdots,\sigma_n|\sigma'_1,\cdots,\sigma'_n)\,,
\eea
where ${\cal A}_n$ represents the full $n$-point amplitude with coupling constants stripped off. The summation over $\sigma$ and $\sigma'$ involves all un-cyclic permutations among the external scalars, encoded by ${\cal S}_n\setminus Z_n$ and ${\cal S}'_n\setminus Z'_n$ respectively. Each partial amplitude ${\cal A}_{\rm BAS}(\sigma_1, \cdots, \sigma_n|\sigma'_1, \cdots, \sigma'_n)$, with color-factor stripped, is simultaneously planar with respect to both orderings. For instance, the first diagram in Figure.\ref{5p} contributes to the $5$-point amplitude ${\cal A}_{\rm BAS}(1,2,3,4,5|1,4,2,3,5)$ as it satisfies both orderings $(1,2,3,4,5)$ and $(1,4,2,3,5)$. Conversely, the second diagram in Figure.\ref{5p} is incompatible with the ordering $(1,4,2,3,5)$ and is therefore excluded. It is straightforward to determine that the first diagram in Figure.\ref{5p} is the only appropriate candidate, thus the amplitude ${\cal A}_{\rm BAS}(1,2,3,4,5|1,4,2,3,5)$ is expressed as follows:
\bea
{\cal A}_{\rm BAS}(1,2,3,4,5|1,4,2,3,5)={1\over s_{23}}{1\over s_{51}}\,,
\eea
up to an overall sign. The Mandelstam variable $s_{i\cdots j}$ is defined as usual
\bea
s_{i\cdots j}\equiv k_{i\cdots j}^2\,,~~~{\rm with}~k_{i\cdots j}\equiv\sum_{a=i}^j\,k_a\,,~~~~\label{mandelstam}
\eea
where $k_a$ is the momentum carried by the external leg $a$.

The anti-symmetry of structure constants $F^{ABC}$ and $f^{abc}$ indicates each partial amplitude carries an overall sign $\pm$,
which arises from swapping lines attached to the same vertex. In this paper, the convention for this overall sign is chosen to be $+$ when the
two orderings identical. For instance, the amplitude ${\cal A}_{\rm BAS}(1,2,3,4|1,2,3,4)$ carries
an overall sign of $+$.
For cases with two different orderings, the overall sign can be determined by counting flips, as detailed in \cite{Cachazo:2013iea}.

We also consider the massless ${\rm YM}\oplus {\rm BAS}$ theory, with the Lagrangian \cite{Chiodaroli:2017ngp}
\bea
{\cal L}_{{\rm YM}\oplus {\rm BAS}}&=&-{1\over4}\,F^{a}_{\mu\nu}\,F^{a\mu\nu}+{1\over2}\,D_\mu\phi^{Aa}\,D^{\mu}\phi^{Aa}
-{g^2\over4}\,f^{abe}\,f^{ecd}\,\phi^{Aa}\phi^{Bb}\phi^{Ac}\phi^{Bd}\nn
& &+{\lambda \over3!}\,F^{ABC}f^{abc}\,
\phi^{Aa}\phi^{Bb}\phi^{Cc}\,.~~\label{lag}
\eea
The indices $a$, $b$, $c$ and $d$ run over the adjoint representation of the gauge group, while scalar
fields carry additional indices
$A$, $B$, $C$. The field strength and covariant derivative are defined as usual
\bea
& &F^{a}_{\mu\nu}=\partial_\mu\,A^a_\nu-\partial_\nu\,A^a_\mu+g\,f^{abc}\,A^b_\mu A^c_\nu\,,\nn
& &D_\mu\phi^{Aa}=\partial_\mu\phi^{Aa}+g\,f^{abc}\,A^b_\mu\phi^{Ac}\,.
\eea

We can decompose the full amplitude without the coupling constant, denoted as ${\cal A}_{n+m}$, according to the trace of generators of the gauge group as follows:
\bea
{\cal A}_{n+m}=\sum_{\sigma\in{\cal S}_{n+m}\setminus Z_{n+m}}\,{\rm tr}[T^{a_{\sigma_1}}\cdots T^{a_{\sigma_{n+m}}}]\,
{\cal A}_{n+m,1}(\sigma_1,\cdots,\sigma_{n+m})\,,~~\label{deco-color}
\eea
where ${\cal S}_{n+m}\setminus Z_{n+m}$ encodes un-cyclic permutations among all external legs including $n$ scalars and $m$ gluons. Meanwhile, decomposing the tree amplitude ${\cal A}_{n+m}$ according to the additional group factor with indices $A,B,C,...$ can be structured as follows
\begin{equation}
\begin{aligned}
&\mathcal{A}_{n+m}\\
=&\sum_{\sigma \in \mathcal{S}_n \backslash Z_n} \operatorname{tr}\left[T^{A_{\sigma_1}} \cdots T^{A_{\sigma_n}}\right] \mathcal{A}_{n+m, 1}\left(\sigma_1, \cdots, \sigma_n\right) \\
+&\sum_{n_1+n_2=n} \sum_{\alpha \in \mathcal{S}_{n_1} \backslash Z_{n_1}} \sum_{\beta \in \mathcal{S}_{n_2} \backslash Z_{n_2}} \operatorname{tr}\left[T^{A_{\alpha_1}} \cdots T^{A_{\alpha_{n_1}}}\right] \operatorname{tr}\left[T^{A_{\beta_1}} \cdots T^{A_{\beta_{n_2}}}\right] \mathcal{A}_{n+m, 2}\left(\left\{\alpha_1, \cdots, \alpha_{n_1}\right\} ::\left\{\beta_1, \cdots, \beta_{n_2}\right\}\right) \\
+&\cdots.
\end{aligned}
\label{multi-trace}
\end{equation}
Each un-cyclic permutation is denoted by \({\cal S}_{n_i} \setminus Z_{n_i}\). The single-trace terms are outlined in the first line of \eqref{multi-trace}. The partial amplitudes $\mathcal{A}_{n+m, 2}\left(\{\alpha_1, \ldots, \alpha_{n_1}\}:: \{\beta_1, \ldots, \beta_{n_2}\}\right)$
are structured into two ordered sets, $\{\alpha_1, \ldots, \alpha_{n_1}\}$ and $\{\beta_1, \ldots, \beta_{n_2}\}$, corresponding to two separate traces. This arrangement of the second line indicates it is the double-trace contributions. The rest omitted portions of \eqref{multi-trace} include additional multi-trace contributions, segmented in a similar manner.

In rest of this paper, we will abbreviate notation ${\cal A}_{\rm BAS}(\sigma_1,\cdots,\sigma_n|\sigma'_1,\cdots,\sigma'_n)$ as ${\cal A}_{\rm BAS}(\vec{\pmb\sigma}_n|\vec{\pmb\sigma}'_n)$, where bold vector notation $\vec{\pmb\sigma}_n$ and $\vec{\pmb\sigma}'_n$ represents two ordered sets each containing $n$ elements while regular font  $\sigma_i$, $\sigma'_i$ denote the $i$-th elements in the corresponding ordered sets. Similarly, we adhere to comparable conventions for YMS amplitudes featuring $n$ external scalars and $m$ external gluons. The single-trace amplitude is represented as
$
{\cal A}_{\rm YMS}(\vec{\pmb\sigma}'_n; \pmb{g}_m|\vec{\pmb\sigma}_{n+m})
$
where $\vec{\pmb\sigma}_{n+m}$ orders all external legs, $\vec{\pmb\sigma}'_n$ comes from the additional group, and $\pmb{g}_m$ denotes the unordered gluons. Double-trace amplitudes are expressed as
$
{\cal A}_{\rm YMS}(\vec{\pmb\rho}_1::\vec{\pmb\rho}_2; \pmb{g}_m|\vec{\pmb\sigma}_{n+m})
$
where $\vec{\pmb\rho}_1$ and $\vec{\pmb\rho}_2$ are ordered sets representing two traces. The format for amplitudes with multi-trace is analogous.

\begin{figure}
  \centering
  \includegraphics[width=10cm]{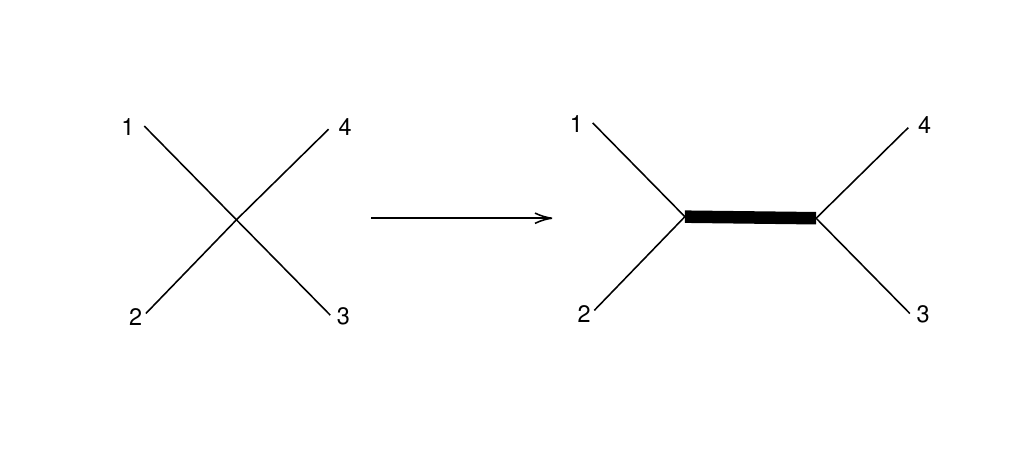} \\
  \caption{Split the $4$-point vertex to $3$-point ones. The bold line corresponds to the inserted propagator $1/s_{12}$.
  This manipulation turns the original numerator $N$ to $s_{12}N$.}\label{1234}
\end{figure}
\begin{figure}
  \centering
  \includegraphics[width=9cm]{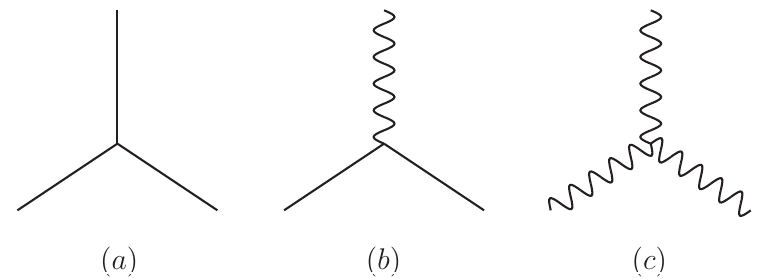} \\
  \caption{Vertices in BAS, YMS and YM theories. Straight lines represent scalars while wave lines represent gluons.}\label{vertex}
\end{figure}

Before concluding this subsection, we wish to highlight that for YM and YMS amplitudes, all interactions can be broken down into trivalent ones. This is because any higher-order vertex can be decomposed into $3$-point vertices by inserting $1=D/D$, where $1/D$ represents propagators connecting the resulting $3$-point vertices. With such insertions, the full amplitude can be represented more succinctly as
\bea
{\cal A}=\sum_{\Gamma}\,{N_\Gamma\over\prod_i D_i}\,,
\eea
where the summation is over all possible diagrams $\Gamma$ with only cubic interactions, and $1/D_i$ represents the corresponding propagators. The additional numerators $D$ created by insertions of $1=D/D$ are absorbed into the numerators $N_\Gamma$. It is worth noting that the decomposing is not unique. An example of splitting a $4$-point vertex into cubic ones is illustrated in Figure \ref{1234}. As a direct consequence of this decomposition, all amplitudes discussed in this subsection involve only three types of interactions, as depicted in Figure \ref{vertex}.

%

\subsection{Soft behaviors of BAS, YMS and YM amplitudes }
\label{subsec-introsoft}

In this part, we undertake a review of the soft behavior exhibited by BAS, YMS, and YM amplitudes.

Given our focus on BAS amplitude, characterized solely by massless propagators, its leading soft behavior emerges exclusively from the divergence of propagators in the soft limit. Here, let $k_i^{\mu}\to \tau k_i^{\mu}$, $\tau\to 0$, with the parameter $\tau$ serving to quantify the degree to an external momentum approaches zero. Accordingly, the leading soft behavior of ${\cal A}_{\rm BAS}(\vec{\pmb\sigma}_n|\vec{{\pmb\sigma}}'_n)$ has the following form \cite{Zhou:2022orv,Du:2024dwm}:
\bea
{\cal A}^{(0)_i}_{\rm BAS}(\vec{\pmb\sigma}_n|\vec{\pmb\sigma}'_n)&=&S^{(0)_i}_s\,
{\cal A}_{\rm BAS}(\vec{\pmb\sigma}_n\setminus i|\vec{\pmb\sigma}'_n\setminus i)\,,~~~\label{soft-s1}
\eea
where the corresponding soft factor of $i$-th external legs reads:
\bea
S^{(0)_i}_s={1\over\tau}\,\sum_{j\neq i}\,{\delta_{ij}\,\delta'_{ij}\over s_{ij}}\,.~~\label{soft-fac-s1}
\eea
In \eref{soft-fac-s1}, the symbol $\delta_{ij}$ is introduced to indicate whether the channel $s_{ij}$ is permitted by two orderings $\vec{\pmb\sigma}_n$ and $\vec{{\pmb\sigma}}'_n$. Put differently, if $i$ and $j$ are not adjacent in $\vec{\pmb\sigma}_n$, then $\delta_{ij}=0$. However, if $i$ and $j$ are adjacent elements, $\delta_{ij}$ takes the value of $1$ when $i$ precedes $j$ and $-1$ when $i$ follows $j$. Similarly, according to ordering $\vec{{\pmb\sigma}}'_n$, the determination of $\delta'_{ij}$ follows identical rules.


%
%
%

%
%
%

The antisymmetry of the symbol $\delta_{ij}=-\delta_{ji}$ results in the following property
\bea
& &\left({\delta_{ab}\over s_{ab}}+{\delta_{be}\over s_{be}}\right)\,
{\cal A}_{\rm BAS}(1,\cdots,a,k_1,\cdots,k_p,e,\cdots,n|\vec{\pmb\sigma}_n\setminus b)\nn
&=&\left({\delta_{ab}\over s_{ab}}+{\delta_{bk_1}\over s_{bk_1}}\right)\,
{\cal A}_{\rm BAS}(1,\cdots,a,k_1,\cdots,k_p,e,\cdots,n|\vec{\pmb\sigma}_n\setminus b)\nn
& &+\sum_{i=1}^{p-1}\,\left({\delta_{k_ib}\over s_{k_ib}}+{\delta_{bk_{i+1}}\over s_{bk_{i+1}}}\right)\,
{\cal A}_{\rm BAS}(1,\cdots,a,k_1,\cdots,k_p,e,\cdots,n|\vec{\pmb\sigma}_n\setminus b)\nn
& &+\left({\delta_{k_pb}\over s_{k_pb}}+{\delta_{be}\over s_{be}}\right)\,
{\cal A}_{\rm BAS}(1,\cdots,a,k_1,\cdots,k_p,e,\cdots,n|\vec{\pmb\sigma}_n\setminus b)\,,~~\label{tech}
\eea
which will be used frequently in subsequent sections.

The same soft theorem for BAS particles is also satisfied by YMS amplitudes, namely,
\bea
{\cal A}^{(0)_i}_{\rm YMS}(1,\cdots,n;\pmb{g}_m|\vec{\pmb\sigma}_{n+m})=S^{(0)_i}_s\,
{\cal A}_{\rm YMS}(1,\cdots,i-1,i+1,\cdots,n;\pmb{g}_m|\vec{\pmb\sigma}_{n+m}\setminus i)\,,~~\label{softs-YMS}
\eea
where the soft factor
$S^{(0)_i}_s$ is the same as that defined in \eref{soft-fac-s1} \cite{Zhou:2022orv,Du:2024dwm}.  It is direct to observe that such leading soft behavior vanishes when the
YMS amplitude includes only two external BAS scalars.

Next, we'll discuss the soft behaviors related to gluons. In the YMS amplitude, let's consider taking gluon $p$ as the soft particle, i.e., letting $k^{\mu}_p\to \tau k^{\mu}_p$ as $\tau\to 0$ with $p\in\pmb{g}_m$. In this scenario, the leading and subleading soft behaviors are \cite{Zhou:2022orv,Du:2024dwm}
\bea
{\cal A}_{\rm YMS}(1,\cdots,n;\pmb{g}_m|\vec{\pmb\sigma}_{n+m})
=\Bigl[S^{(0)_{p}}_g+S^{(1)_{p}}_g\Bigr]\,{\cal A}_{\rm YMS}(1,\cdots,n;\pmb{g}_m\setminus p|\vec{\pmb\sigma}_{n+m}\setminus p)
+{\cal O}(\tau)\,,~~\label{softg-YMS}
\eea
where the leading and subleading soft factors are respectively given as
\bea
S^{(0)_{p}}_g&=&{1\over\tau}\,\sum_{a\in\{1,\cdots,n\}\cup\{g_k\}\setminus p}\,{\delta_{ap}\,(\epsilon_{p}\cdot k_a)\over s_{ap}}\,,\nn
S^{(1)_{p}}_g&=&\sum_{a\in\{1,\cdots,n\}\cup\{g_k\}\setminus p}\,{\delta_{ap}\,
\big(\epsilon_{p}\cdot J_a\cdot k_{p}\big)\over s_{ap}}\,.~~~\label{soft-fac-g-0-2}
\eea
In the above, $\delta_{ap}$ is determined by the ordering of $\vec{\pmb\sigma}_{n+m}$. $\epsilon_a$ represents the polarization vector of external gluon $a$, while $J_b$ denotes the angular momentum of external particle $b$, without differentiation between scalars or gluons. It's noteworthy that the aforementioned soft theorem extends to pure YM amplitudes \cite{Casali:2014xpa},
\bea
{\cal A}_{\rm YM}(\vec{\pmb\sigma}_n)=\Bigl[S^{(0)_{p}}_g+S^{(1)_{p}}_g\Bigr]\,{\cal A}_{\rm YM}(\vec{\pmb\sigma}_n\setminus p)
+{\cal O}(\tau)\,,~~\label{soft-YM}
\eea
with the same soft factors $S^{(0)_{p}}_g$ and $S^{(1)_{p}}_g$ given in \eref{soft-fac-g-0-2}.
In the subleading soft operator $S^{(1)_{p}}_g$ in \eref{soft-fac-g-0-2}, the angular momentum $J_a^{\mu\nu}$ acts
on $k^\rho_a$  and on $\epsilon^\rho_a$ with the orbital and the spin parts of the generator
\bea
J_a^{\mu\nu}\,k_a^\rho= k_a^{\mu}\,
{\partial k_a^\rho\over\partial k_{a,\nu}}-k_a^{\nu}\,{\partial k_a^\rho\over\partial k_{a,\mu}}\,,~~~~
J_a^{\mu\nu}\,\epsilon_a^\rho=\big(\eta^{\nu\rho}\,\delta^\mu_\sigma-\eta^{\mu\rho}\,\delta^\nu_\sigma\big)\,
\epsilon^\sigma_a\,.~~\label{pro}
\eea
Using \eref{pro}, we can give the following useful relations
\bea
\big(S^{(1)_p}_g\,V_a\big)\cdot V_b={\delta_{pa}\over s_{pa}}\,(V_a\cdot f_p\cdot V_b)\,,~~~~
V_a\cdot\big(S^{(1)_p}_g\,V_b\big)={\delta_{ap}\over s_{ap}}\,(V_a\cdot f_p\cdot V_b)\,,~~~~\label{iden-1}
\eea
and
\bea
V_1\cdot\big(S^{(1)_p}_g\,f_a\big)\cdot V_2={\delta_{ap}\over s_{ap}}\,V_1\cdot(f_p\cdot f_a-f_a\cdot f_p)\cdot V_2\,,~~~~\label{iden-2}
\eea
The strength tensor $f^{\mu\nu}_p$ is defined as $f^{\mu\nu}_p\equiv k^\mu_p\epsilon^\nu_p-\epsilon^\mu_p k^\nu_p$. $V_a$ and $V_b$ in \eref{iden-1} represent the external momentum or polarization vectors for the corresponding external legs $a$ or $b$, while $V_1$ and $V_2$ In \eref{iden-2} can be chosen as any Lorentz vectors.

To effectively apply soft theorems to the new cases addressed in this paper, it is pertinent to discuss the universality of the aforementioned soft behaviors. It is a natural expectation that, within the same theory, the soft behavior remains universal for particles of the same type, irrespective of the number of external legs. For example the soft behavior of scalars in YMS theories. However, when comparing two different theories, it is generally unreasonable to assume the universality of soft behavior. For instance, one should allow for modifications to the soft theorem for YM amplitudes when introducing new types of interactions between gluons. We will now elucidate that the universality of soft behavior for scalars in BAS and YMS theories, as well as the universality of soft behavior for gluons in YMS and YM theories, are based on specific underlying reasons.

\begin{figure}
  \centering
  \includegraphics[width=5cm]{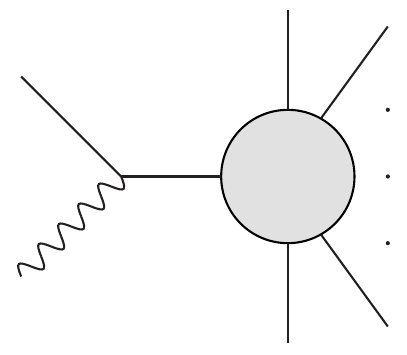} \\
  \caption{Diagram which has no contribution to the leading soft behavior of an external scalar.}\label{exclu}
\end{figure}

In BAS, YMS, and pure YM theories, there exist only three types of interactions, shown in Figure \ref{vertex}, as any higher-point vertex can be decomposed into trivalent ones. The leading soft theorem for BAS amplitudes applies to the YMS case due to the leading soft behavior, where an external scalar in a YMS amplitude receives contributions solely from the type $(a)$ interactions in Figure \ref{vertex}. This assertion can be verified using the corresponding soft factor in \eref{soft-fac-s1}, where $\delta_{ij}\delta'_{ij}$ does not affect any external gluon, as each gluon appears in only one ordering. This observation rules out contributions from diagrams as depicted in Figure \ref{exclu}, leaving only contributions from the type (a) vertices in Figure \ref{vertex}. The exclusion of contributions from Figure \ref{exclu} can also be comprehended by analyzing the diagram itself. In the soft limit, the soft particle can be regarded as vanishing, hence removing the soft scalar leaves us with a diagram representing a gluon being converted into a scalar without any interaction. Such a process is physically untenable, thus the corresponding diagram does not contribute to the soft behavior.

\begin{figure}
  \centering
  \includegraphics[width=9cm]{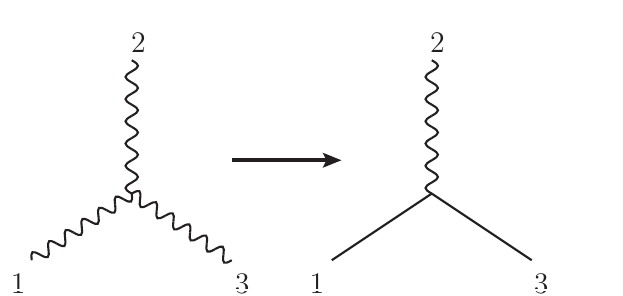} \\
  \caption{Generating $3$-point YMS amplitude from the YM one.}\label{DR1}
\end{figure}

For gluon cases, a YMS amplitude involves both type (b) and type (c) vertices from Figure \ref{vertex}, while a pure YM amplitude only features type (c) vertices. However, both YMS and YM amplitudes share the same soft theorem for external gluons because of a key property: a $3$-point on-shell YMS amplitude with a type (b) interaction can be generated from a $3$-point on-shell YM amplitude with a type (c) interaction through dimensional reduction, as depicted in Figure \ref{DR1}. To elaborate, in a $(d+1)$ dimensional space-time setting, we position $\epsilon_1$ and $\epsilon_3$ in the extra dimension, while $\epsilon_2$ and all external momenta remain in the usual $d$-dimensional space-time. This setup makes external gluons $1$ and $3$ behave like two scalars from a $d$-dimensional viewpoint. While generally, this relationship doesn't imply equivalence between type (b) and type (c) vertices due to potential off-shell contributions, in the soft limit, however, each internal off-shell line sharing the same trivalent vertex with the soft gluon becomes on-shell. Consequently, from the $(d+1)$ dimensional perspective, contributions from both type (b) and type (c) vertices are equivalent in the soft limit.

In this paper, we explore novel interactions between gluons with mass dimensions higher than those in YM theory. We affirm that the soft theorem for BAS scalars remains applicable when these scalars are coupled to gluons. The soft operator in \eref{soft-fac-s1} consistently prohibits contributions from external gluons. However, when considering an external gluon as the soft particle, it's logical to partition the full amplitude into two components: 1) The soft gluon attached to an ordinary vertex in YMS or YM theory follows the same soft theorem described in \eref{softg-YMS}, \eref{soft-YM} and \eref{soft-fac-g-0-2}; 2) the soft gluon interacts with a new vertex possessing a higher mass dimension. In next section, we will argue and validate that contributions from this second type of interaction vanish at both leading and subleading orders.

\section{Tree amplitudes of YM$^{+2}$}
\label{sec-YM+2}

The amplitudes which will be constructed in this section are those described by the Lagrangian
\bea
{\cal L}_{\rm YM^{+2}}=-{1\over4}\,F^{a}_{\mu\nu}F^{a\mu\nu}-{g\over 3\Lambda^2}\,f^{abc}F^{a~\nu}_{~\mu}F^{b~\rho}_{~\nu}F^{c~\mu}_{~\rho}\,,~~\label{Lag-+2}
\eea
and we restrict our selves to the amplitudes with the single insertion of local operator $F^3\equiv f^{abc}F^{a~\nu}_{~\mu}F^{b~\rho}_{~\nu}F^{c~\mu}_{~\rho}$. The corresponding Cachazo-He-Yuan (CHY) formula for such amplitudes was constructed in \cite{He:2016iqi}. Our method is based on exploiting the subleading soft theorem for YM amplitudes, without the aid of above Lagrangian or CHY formula.

In section \ref{subsec-YM+2-3p}, we bootstrap the $3$-point YM amplitudes with mass dimension ${\cal D}+2$. Section \ref{subsec-softbehavior} discusses the soft behavior at the subleading order under consideration. In sections \ref{subsec-YM+2-4p} and \ref{subsec-YM+2-np}, we recursively construct higher-point YM amplitudes with mass dimension ${\cal D}+2$, starting from the $3$-point ones determined in section \ref{subsec-YM+2-3p}.

\subsection{3-point amplitudes}
\label{subsec-YM+2-3p}

For the standard YM theory, the color ordered $3$-point amplitude ${\cal A}_{\rm YM}(1,2,3)$ can be fixed as
\bea
{\cal A}_{\rm YM}(1,2,3)=C\,(\epsilon_1\cdot\epsilon_2)\,(\epsilon_3\cdot k_1)+{\rm cyclic}\,,~~\label{YM-3p}
\eea
due to the Lorentz invariance, the linear dependence on each polarization, as well as the requirement of mass dimension. The momentum conservation and the on-shell condition $\epsilon_i\cdot k_i=0$ ensure that replacing $k_1$ in \eref{YM-3p} by $k_2$ leads to the equivalent expression, up to an overall sign. Now, let us consider whether different $3$-point ordered amplitudes can be constructed without requiring the mass dimension to be $1$. The unique answer is:
\bea
{\cal A}_{\rm YM^{+2}}(1,2,3)=C'\,(\epsilon_1\cdot k_2)\,(\epsilon_2\cdot k_3)\,(\epsilon_3\cdot k_1)\,,~~\label{3p-YM+2}
\eea
according to the on-shell condition $k_i\cdot k_j=0$ for any $i,j\in\{1,2,3\}$ in the $3$-point case. Here we use the subscript ${\rm YM^{+2}}$
to denote amplitudes of gluons with the mass dimension ${\cal D}+2$, where ${\cal D}$ stands for the mass dimension of standard YM amplitudes. As can be verified, \eref{3p-YM+2} is the ordered amplitude for gluons with the $F^3$ interaction, corresponds to the Lagrangian in \eref{Lag-+2}. The amplitude ${\cal A}_{\rm YM^{+2}}(3,2,1)$ can be generated from \eref{3p-YM+2} via the replacement $1\leftrightarrow3$ which is equivalent to adding an overall $-$. Since $1,2,3$ and $3,2,1$ are the only two possible orderings for the $3$-point case, all $3$-point amplitudes with mass dimension ${\cal D}+2$ have been determined.

To perform the recursive construction based on soft behaviors, we rewrite ${\cal A}_{\rm YM^{+2}}(\vec{\pmb{\sigma}}_3)$ as the following combination of single-trace YMS and pure BAS amplitudes,
\bea
{\cal A}_{\rm YM^{+2}}(\vec{\pmb{\sigma}}_3)&=&\alpha\,\Big({\rm tr}(f_2\cdot f_1)\,{\cal A}_{\rm YMS}(1,2;3|\vec{\pmb{\sigma}}_3)
+{\rm cyclic}\Big)\nn
& &+\beta\,\Big({\rm tr}(f_3\cdot f_2\cdot f_1)\,{\cal A}_{\rm BAS}(1,2,3|\vec{\pmb{\sigma}}_3)+{\rm tr}(f_1\cdot f_2\cdot f_3)\,
{\cal A}_{\rm BAS}(3,2,1|\vec{\pmb{\sigma}}_3)\Big)\,,~~\label{expan-3p-YM+2}
\eea
where $\vec{\pmb{\sigma}}_3$ denotes the general ordering among three external legs. We have used the expansion \cite{Fu:2017uzt}
\bea
{\cal A}_{\rm YMS}(1,2;3|\vec{\pmb{\sigma}}_3)=(\epsilon_3\cdot k_1)\,{\cal A}_{\rm BAS}(1,3,2|\vec{\pmb{\sigma}}_3)\,,
\eea
to get the expansion formula \eref{expan-3p-YM+2}. Each coefficient ${\rm tr}(f_{a_1}\cdot f_{a_2}\cdots f_{a_\ell})$ is defined as
\bea
{\rm tr}(f_{a_1}\cdot f_{a_2}\cdots f_{a_\ell})=(f_{a_1})_{\mu_1}^{~\mu_2}\,(f_{a_2})_{\mu_2}^{~\mu_3}\cdots(f_{a_\ell})_{\mu_\ell}^{~\mu_1}\,.
\eea
For the current $3$-point case, this formula includes some vanishing terms since $k_i\cdot k_j=0$, but it manifests gauge invariance for each polarization due to the definition of the strength tensor $f_i^{\mu\nu}$. We do not fix parameters $\alpha$ and $\beta$ at this step, since the first and second lines on the right-hand side of \eref{expan-3p-YM+2} are equivalent to each other. Because of the different ways of inserting vanishing terms, the two lines do not result in the same $4$-point object when inverting the soft theorem, as will be seen in section \ref{subsec-YM+2-4p}. Thus, parameters $\alpha$ and $\beta$ will be fixed in section \ref{subsec-YM+2-4p} by considering the consistent $4$-point amplitudes.

\subsection{Soft behavior at subleading order}
\label{subsec-softbehavior}

\begin{figure}
  \centering
  \includegraphics[width=12cm]{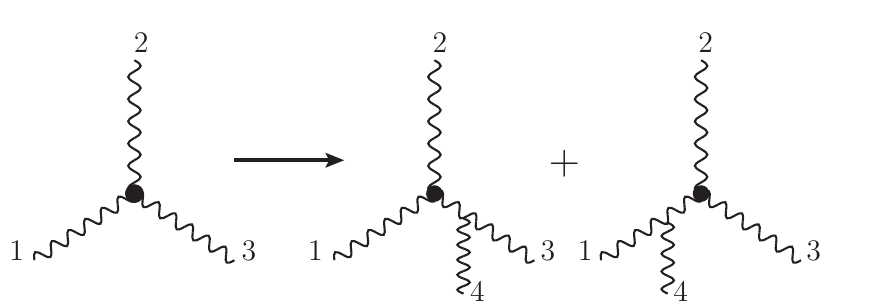} \\
  \caption{Inserting the leg $4$ into the $3$-point ${\rm YM}^{+2}$ amplitude.
  The big bold point represents the ${\rm YM}^{+2}$ vertex.}\label{adding}
\end{figure}

In this subsection, we briefly discuss our method, focusing particularly on the soft behavior at the $\tau^0$ order. As explained in section \ref{subsec-introBASYMS}, all higher-point interactions can be broken down into cubic ones, so we solely concentrate on trivalent vertices. To clarify, we denote the interaction vertices in ordinary YM theory as YM vertices, and those carrying a mass dimension higher than the ordinary ones by $2$ as ${\rm YM}^{+2}$ vertices. Each ${\rm YM}^{+2}$ amplitude ${\cal A}_{\rm YM^{+2}}(\vec{\pmb{\sigma}}_n)$ contains only one cubic ${\rm YM}^{+2}$ vertex to maintain the correct mass dimension. Thus, inserting an additional external gluon into ${\cal A}_{\rm YM^{+2}}(\vec{\pmb{\sigma}}_n)$ via a YM vertex results in a subpart of the full amplitude ${\cal A}{\rm YM^{+2}}(\vec{\pmb{\sigma}}_{n+1})$. Feynman diagrams illustrating such insertion, with $\vec{\pmb{\sigma}}_3=\{1,2,3\}$ and $\vec{\pmb{\sigma}}_4=\{1,2,3,4\}$, are shown in Figure \ref{adding}. Both resulting diagrams contribute to ${\cal A}_{\rm YM^{+2}}(\vec{\pmb{\sigma}}_4)$.

We implement such insertions using the subleading soft theorem for YM amplitudes, which describes the soft behavior for $k_i\to\tau k_i$, $\tau\to0$, at the $\tau^0$ order. Any diagram where the soft gluon $i$ is connected to a ${\rm YM}^{+2}$ vertex contributes nothing at the $\tau^0$ order. If $i$ and $j$ are connected to a common ${\rm YM}^{+2}$ vertex, then the propagator $1/s_{ij}$ becomes on-shell as $\tau\to0$, causing the full $n$-point amplitude to factorize into a $3$-point on-shell ${\rm YM}^{+2}$ amplitude and another $(n-1)$-point one. The leading order of the on-shell $3$-point ${\rm YM}^{+2}$ amplitude is the $\tau^2$ order, determined by the expression in \eref{3p-YM+2}, while the leading orders of the propagator $1/s_{ij}$ and the $(n-1)$-point amplitude are $\tau^{-1}$ and $\tau^0$, respectively. Hence, the leading order of contributions from such diagrams is the $\tau^1$ order.

Thus, the only contribution at the $\tau^0$ order comes from diagrams where $i$ is connected to ordinary YM vertices, as depicted in Figure \ref{adding}. This observation implies that ${\cal A}^{(1)i}_{\rm YM^{+2}}(\vec{\pmb\sigma}_n)$ satisfies the standard subleading soft theorem for YM amplitudes, namely:
\bea
{\cal A}^{(1)_i}_{\rm YM^{+2}}(\vec{\pmb\sigma}_n)=S^{(1)_i}_g\,{\cal A}_{\rm YM^{+2}}(\vec{\pmb\sigma}_{n-1})\,,
~~\label{softbehav}
\eea
with the soft factor $S^{(1)_i}_g$ given in \eref{soft-fac-g-0-2}. The reason we opt for the subleading soft behavior over the leading one is as follows. When the soft gluon is attached to a YM vertex, the on-shell propagator leads to the factorization of the amplitude into an on-shell $3$-point YM amplitude and another $(n-1)$-point ${\rm YM}^{+2}$ one. Then by \eref{YM-3p}, we find that the soft behavior of the $3$-point on-shell YM amplitude can be split into two parts at different orders, $\tau^0$ and $\tau^1$. For example, take the gluon $3$ to be soft, then the term $(\epsilon_1\cdot\epsilon_2)(\epsilon_3\cdot k_1)$ in \eref{YM-3p} is at the $\tau^0$ order, while the terms $(\epsilon_2\cdot\epsilon_3)(\epsilon_1\cdot k_2)$ and $(\epsilon_3\cdot\epsilon_1)(\epsilon_2\cdot k_3)$ are at the $\tau^1$ order, due to the on-shell condition $\epsilon_i\cdot k_i=0$ and momentum conservation. Combining this with the propagator $1/s_{ij}$, which carries $\tau^{-1}$, and the $(n-1)$-point amplitude whose leading order is $\tau^0$, we see that the two $\tau^1$ order $3$-point terms never contribute to the $\tau^{-1}$ order of the full amplitude.
On the other hand, the soft behavior of $\tau^0$ order contains all the information when the soft gluon links to a YM vertex.

\begin{figure}
  \centering
  \includegraphics[width=8cm]{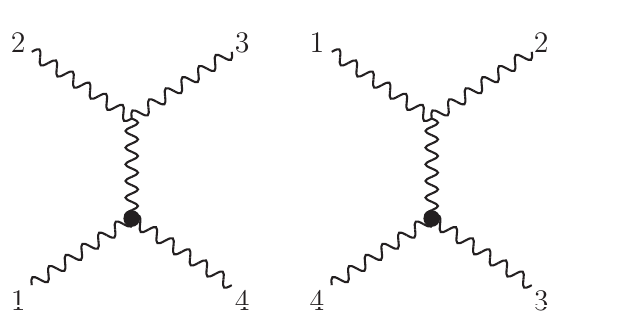} \\
  \caption{Diagrams correspond to missed terms.}\label{missing}
\end{figure}

The undetectable terms at the $\tau^0$ order are those from diagrams where $i$ is connected to ${\rm YM}^{+2}$ vertices will be revealed by a cyclic permutation of external legs. Because each $n$-point trivalent diagram includes at least one YM vertex for $n\geq4$, there always exists an external leg $i$ such that under the soft limit $k_i\to\tau k_i$, the contributions from these diagrams are detectable at the $\tau^0$ order. For instance, while the two diagrams in Figure \ref{missing} are undetectable for $k_4\to\tau k_4$, they become detectable for $k_2\to\tau k_2$. Thus, by enforcing cyclicity, these missed terms will be reinstated. Further discussions on this topic will be presented in section \ref{subsec-YM+2-4p}.

At the end of this subsection, it is worth noting that although $4$-point diagrams in Figure \ref{adding} and \ref{missing} are related by cyclic permutation,
we do not require the contributions form them to inherit such symmetry. This is due to the fact that we have not fixed the off-shell form of any trivalent vertex. In particular, as mentioned earlier, all $4$-point vertices are considered as combinations of $3$-point ones.  A $4$-point ${\rm YM}^{+2}$ vertex can be decomposed into a cubic ${\rm YM}^{+2}$ vertex and a cubic YM vertex, allowing it to be detectable. However, this decomposition is not unique. The flexibility in decomposition breaks the cyclic symmetry from the diagrams in Figure \ref{adding} and \ref{missing}.

\subsection{$4$-point amplitudes}
\label{subsec-YM+2-4p}

Now we attempt to construct $4$-point ${\rm YM}^{+2}$
amplitudes ${\cal A}_{\rm YM^{+2}}(\vec{\pmb{\sigma}}_4)$ from $3$-point ${\cal A}_{\rm YM^{+2}}(\vec{\pmb{\sigma}}_3)$. Consider $k_4\to\tau k_4$, $\tau\to0$. The subleading soft behavior of ${\cal A}_{\rm YM^{+2}}(\vec{\pmb{\sigma}}_4)$ reads
\bea
{\cal A}^{(1)_4}_{\rm YM^{+2}}(\vec{\pmb{\sigma}}_4)&=&S^{(1)_4}_g\,{\cal A}_{\rm YM^{+2}}(\vec{\pmb{\sigma}}_4\setminus4)\nn
&=&S^{(1)_4}_g\,\Big(\alpha\,\big[{\rm tr}(f_2\cdot f_1)\,{\cal A}_{\rm YMS}(1,2;3|\vec{\pmb{\sigma}}_4\setminus4)+{\rm cyclic}\big]\nn
& &+\beta\,\big[{\rm tr}(f_3\cdot f_2\cdot f_1)\,{\cal A}_{\rm BAS}(1,2,3|\vec{\pmb{\sigma}}_4\setminus4)
+{\rm tr}(f_1\cdot f_2\cdot f_3)\,{\cal A}_{\rm BAS}(3,2,1|\vec{\pmb{\sigma}}_4\setminus4)\big]\Big)\,,
\eea
where the second equality is obtained by substituting the expansion of $3$-point amplitudes in \eref{expan-3p-YM+2}.
One can evaluate the first part as
\bea
& &S^{(1)_4}_g\,\Big({\rm tr}(f_2\cdot f_1)\,{\cal A}_{\rm YMS}(1,2;3|\vec{\pmb{\sigma}}_4\setminus4)+{\rm cyclic}\Big)\nn
&=&{\rm tr}(f_2\cdot f_1)\,\Big(S^{(1)_4}_g\,{\cal A}_{\rm YMS}(1,2;3|\vec{\pmb{\sigma}}_4\setminus4)\Big)+\Big(S^{(1)_4}_g\,
{\rm tr}(f_2\cdot f_1)\Big)\,{\cal A}_{\rm YMS}(1,2;3|\vec{\pmb{\sigma}}_4\setminus4)+{\rm cyclic}\nn
&=&{\rm tr}(f_2\cdot f_1)\,{\cal A}^{(1)_4}_{\rm YMS}(1,2;\{3,4\}|\vec{\pmb{\sigma}}_4)\nn
& &+\tau\,{\rm tr}(f_4\cdot f_2\cdot f_1)\,{\cal A}^{(0)_4}_{\rm YMS}(1,2,4;3|\vec{\pmb{\sigma}}_4)
+\tau\,{\rm tr}(f_2\cdot f_4\cdot f_1)\,{\cal A}^{(0)_4}_{\rm YMS}(1,4,2;3|\vec{\pmb{\sigma}}_4)\nn
& &+{\rm cyclic}\,.~~\label{4p-p1}
\eea
In the above, the first equality arises from the Leibnitz rule since $S^{(1)_4}_g$ is a differential operator.
The second equality uses the soft theorems \eref{soft-s1}, \eref{softg-YMS}, i.e.,
\bea
{\cal A}^{(1)_4}_{\rm YMS}(1,2;\{3,4\}|\vec{\pmb{\sigma}}_4)&=&S^{(1)_4}_g\,{\cal A}_{\rm YMS}(1,2;3|\vec{\pmb{\sigma}}_4\setminus4)\,,\nn
{\cal A}^{(0)_4}_{\rm YMS}(\vec{\pmb{\sigma}}'_3;3|\vec{\pmb{\sigma}}_4)&=&{1\over\tau}\,\sum_{i\in\{1,2\}}\,{\delta'_{i4}\,
\delta_{i4}\over s_{i4}}\,{\cal A}_{\rm YMS}(\vec{\pmb{\sigma}}'_3\setminus4;3|\vec{\pmb{\sigma}}_4\setminus4)\,,~~\label{soft-4p}
\eea
as well as the relation in \eref{iden-2}. In \eref{soft-4p}, $\vec{\pmb{\sigma}}'_3$ stands for the ordering among external legs in $\{1,2,4\}$.
Notice that the cyclic terms in the last line of \eref{4p-p1} are generated by the cyclic permutation $1\to2\to3\to1$ without including the leg $4$.

The similar calculation gives
\bea
& &S^{(1)_4}_g\,\Big({\rm tr}(f_3\cdot f_2\cdot f_1)\,{\cal A}_{\rm YMS}(1,2,3|\vec{\pmb{\sigma}}_4\setminus4)\Big)\nn
&=&{\rm tr}(f_3\cdot f_2\cdot f_1)\,{\cal A}^{(1)_4}_{\rm YMS}(1,2,3;4|\vec{\pmb{\sigma}}_4)\nn
& &+\tau\,{\rm tr}(f_4\cdot f_3\cdot f_2\cdot f_1)\,{\cal A}^{(0)_4}_{\rm BAS}(1,2,3,4|\vec{\pmb{\sigma}}_4)
+\tau\,{\rm tr}(f_3\cdot f_4\cdot f_2\cdot f_1)\,{\cal A}^{(0)_4}_{\rm BAS}(1,2,4,3|\vec{\pmb{\sigma}}_4)\nn
& &+\tau\,{\rm tr}(f_3\cdot f_2\cdot f_4\cdot f_1)\,{\cal A}^{(0)_4}_{\rm BAS}(1,4,2,3|\vec{\pmb{\sigma}}_4)\,,~~\label{4p-p2}
\eea
and
\bea
& &S^{(1)_4}_g\,\Big({\rm tr}(f_1\cdot f_2\cdot f_3)\,{\cal A}_{\rm YMS}(3,2,1|\vec{\pmb{\sigma}}_4\setminus4)\Big)\nn
&=&S^{(1)_4}_g\,\Big({\rm tr}(f_1\cdot f_2\cdot f_3)\,{\cal A}_{\rm YMS}(1,3,2|\vec{\pmb{\sigma}}_4\setminus4)\Big)\nn
&=&{\rm tr}(f_2\cdot f_3\cdot f_1)\,{\cal A}^{(1)_4}_{\rm YMS}(1,3,2;4|\vec{\pmb{\sigma}}_4)\nn
& &+\tau\,{\rm tr}(f_4\cdot f_2\cdot f_3\cdot f_1)\,{\cal A}^{(0)_4}_{\rm BAS}(1,3,2,4|\vec{\pmb{\sigma}}_4)
+\tau\,{\rm tr}(f_2\cdot f_4\cdot f_3\cdot f_1)\,{\cal A}^{(0)_4}_{\rm BAS}(1,3,4,2|\vec{\pmb{\sigma}}_4)\nn
& &+\tau\,{\rm tr}(f_2\cdot f_3\cdot f_4\cdot f_1)\,{\cal A}^{(0)_4}_{\rm BAS}(1,4,3,2|\vec{\pmb{\sigma}}_4)\,.~~\label{4p-p3}
\eea
In the first equality of \eref{4p-p3}, we used the cyclic invariance of trace and ordering to fix the leg $1$ at the first
position of the ordering $1,3,2$, this is the convention which will be used in \eref{soft-4p-resul} and \eref{4p-part}.
Substituting \eref{4p-p1}, \eref{4p-p2} and \eref{4p-p3} into \eref{soft-4p}, we obtain
\bea
{\cal A}^{(1)_4}_{\rm YM^{+2}}(\vec{\pmb{\sigma}}_4)&=&\alpha\,\sum_{\underline{i},j\in\{1,2,3\}}\,{\rm tr}(f_j\cdot f_i)\,
{\cal A}^{(1)_4}_{\rm YMS}(i,j;\{1,2,3,4\}\setminus\{i,j\}|\vec{\pmb{\sigma}}_4)\nn
& &+\alpha\,\tau\,\sum_{\underline{i},j\in\{1,2,3\}}\,{\rm tr}(f_4\cdot f_j\cdot f_i)\,
{\cal A}^{(0)_4}_{\rm YMS}(i,j,4;\{1,2,3\}\setminus\{i,j\}|\vec{\pmb{\sigma}}_4)\nn
& &+\alpha\,\tau\,\sum_{\underline{i},j\in\{1,2,3\}}\,{\rm tr}(f_j\cdot f_4\cdot f_i)\,
{\cal A}^{(0)_4}_{\rm YMS}(i,4,j;\{1,2,3\}\setminus\{i,j\}|\vec{\pmb{\sigma}}_4)\nn
& &+\beta\,\sum_{\underline{a},b,c\in\{1,2,3\}}\,{\rm tr}(f_c\cdot f_b\cdot f_a)\,
{\cal A}^{(1)_4}_{\rm YMS}(a,b,c;4|\vec{\pmb{\sigma}}_4)\nn
& &+\beta\,\tau\,\sum_{\underline{a},b,c\in\{1,2,3\}}\,{\rm tr}(f_4\cdot f_c\cdot f_b\cdot f_a)\,
{\cal A}^{(0)_4}_{\rm BAS}(a,b,c,4|\vec{\pmb{\sigma}}_4)\nn
& &+\beta\,\tau\,\sum_{\underline{a},b,c\in\{1,2,3\}}\,{\rm tr}(f_c\cdot f_4\cdot f_b\cdot f_a)\,
{\cal A}^{(0)_4}_{\rm BAS}(a,b,4,c|\vec{\pmb{\sigma}}_4)\nn
& &+\beta\,\tau\,\sum_{\underline{a},b,c\in\{1,2,3\}}\,{\rm tr}(f_c\cdot f_b\cdot f_4\cdot f_a)\,
{\cal A}^{(0)_4}_{\rm BAS}(a,4,b,c|\vec{\pmb{\sigma}}_4)\,,~~~\label{soft-4p-resul}
\eea
which indicates
\bea
{\cal P}^4_{\rm YM^{+2}}(\vec{\pmb{\sigma}}_4)&=&\alpha\,\sum_{\underline{i},j\in\{1,2,3\}}\,{\rm tr}(f_j\cdot f_i)\,
{\cal A}_{\rm YMS}(i,j;\{1,2,3,4\}\setminus\{i,j\}|\vec{\pmb{\sigma}}_4)\nn
& &+\alpha\,\sum_{\underline{i},j\in\{1,2,3\}}\,{\rm tr}(f_4\cdot f_j\cdot f_i)\,
{\cal A}_{\rm YMS}(i,j,4;\{1,2,3\}\setminus\{i,j\}|\vec{\pmb{\sigma}}_4)\nn
& &+\alpha\,\sum_{\underline{i},j\in\{1,2,3\}}\,{\rm tr}(f_j\cdot f_4\cdot f_i)\,
{\cal A}_{\rm YMS}(i,4,j;\{1,2,3\}\setminus\{i,j\}|\vec{\pmb{\sigma}}_4)\nn
& &+\beta\,\sum_{\underline{a},b,c\in\{1,2,3\}}\,{\rm tr}(f_c\cdot f_b\cdot f_a)\,
{\cal A}_{\rm YMS}(a,b,c;4|\vec{\pmb{\sigma}}_4)\nn
& &+\beta\,\sum_{\vec{\pmb{\sigma}}'_4}\,{\rm tr}(F_{\vec{\pmb{\sigma}}'_4})\,
{\cal A}_{\rm BAS}(\vec{\pmb{\sigma}}'_4|\vec{\pmb{\sigma}}_4)\,,~~\label{4p-part}
\eea
where the notation ${\cal P}^i_{\rm YM^{+2}}(\vec{\pmb{\sigma}}_4)$ was introduced
to emphasize this is the detectable part when considering $k_i\to\tau k_i$, rather than the full amplitude.
Here $\underline{i}$ means $i<j$, and similarly $\underline{a}$ means $a<b$, $a<c$.
These constraints serve as gauge fixings for cyclic permutations, which avoid the
over counting of equivalent orderings in summations. In the last line of \eref{4p-part},
$\vec{\pmb{\sigma}}'_4$ stands for the ordered set of legs in $\{1,2,3,4\}$, and the summation
is for all inequivalent $\vec{\pmb{\sigma}}'_4$. The tensor $F^{\mu\nu}_{\vec{\pmb{\sigma}}'_4}$ is defined as
\bea
F^{\mu\nu}_{\vec{\pmb{\sigma}}'_4}=(f_{\sigma_4}\cdot f_{\sigma_3}\cdot f_{\sigma_2}\cdot f_{\sigma_1})^{\mu\nu}\,,
~~~~{\rm for}~\vec{\pmb{\sigma}}'_4=\{\sigma_1,\sigma_2,\sigma_3,\sigma_4\}\,,
\eea
and ${\rm tr}(F_{\vec{\pmb{\sigma}}'_4})$ is understood as $(F_{\vec{\pmb{\sigma}}'_4})_\mu^{~\mu}$.

The complete amplitude ${\cal A}_{\rm YM^{+2}}(\vec{\pmb{\sigma}}_4)$ can be obtained by imposing
a very nature requirement that the amplitude is invariant under the cyclic permutation of external legs,
which is just the basic character of ordering $\vec{\pmb\sigma}_4$. It is equivalent to state that the expansion
of ${\cal A}_{\rm YM^{+2}}(\vec{\pmb{\sigma}}_4)$ to YMS and BAS amplitudes should be invariant
if we perform the cyclic permutation $1\to2\to3\to4\to1$ while keeping $\vec{\pmb{\sigma}}_4$ unaltered.
This condition requires $\alpha=\beta$, and forces us to add three new terms
${\rm tr}(f_4\cdot f_2){\cal A}_{\rm YMS}(2,4;1,3|\vec{\pmb{\sigma}}_4)$,
${\rm tr}(f_4\cdot f_3){\cal A}_{\rm YMS}(3,4;1,2|\vec{\pmb{\sigma}}_4)$ and
${\rm tr}(f_4\cdot f_1){\cal A}_{\rm YMS}(1,4;2,3|\vec{\pmb{\sigma}}_4)$ to \eref{4p-part}. Thus we can chose $\alpha=\beta=1$ to get
\bea
{\cal A}_{\rm YM^{+2}}(\vec{\pmb{\sigma}}_4)=\sum_{\vec{\pmb{\rho}}|\pmb\rho|\geq2}\,{\rm tr}(F_{\vec{\pmb{\rho}}})\,
{\cal A}_{\rm YMS}(\vec{\pmb{\rho}};\{1,2,3,4\}\setminus\pmb\rho|\vec{\pmb{\sigma}}_4)\,.~~~\label{4p-amp}
\eea
Here $\pmb\rho$ is a subset of $\{1,2,3,4\}$ with at least two elements (denoted as $|\pmb\rho|\geq2$), and $\vec{\pmb{\rho}}$
is an ordered set obtained by giving an order to elements in $\pmb\rho$. The tensor $F^{\mu\nu}_{\vec{\pmb{\rho}}}$ is given by
\bea
F^{\mu\nu}_{\vec{\pmb{\rho}}}=(f_{\rho_k}\cdots f_{\rho_1})^{\mu\nu}\,,~~~~{\rm for}~\vec{\pmb{\rho}}=\{\rho_1,\cdots,\rho_k\}\,.
\eea
One can formally extend the summation to all inequivalent $\vec{\pmb{\rho}}$,
since the antisymmetry of the strength tensor $f_i^{\mu\nu}$ indicates that
${\rm tr}(F_{\vec{\pmb{\rho}}})$ vanishes if $\vec{\pmb{\rho}}$ includes less than two element.

Some remarks are in order. First, in the previous section \ref{subsec-softbehavior}, we argued that the undetectable terms for $k_4\to\tau k_4$ vanish at the $\tau^0$ order. The vanishing of missed terms ${\rm tr}(f_4\cdot f_2){\cal A}_{\rm YMS}(2,4;1,3|\vec{\pmb{\sigma}}_4)$, ${\rm tr}(f_4\cdot f_3){\cal A}_{\rm YMS}(3,4;1,2|\vec{\pmb{\sigma}}_4)$, and ${\rm tr}(f_4\cdot f_1){\cal A}_{\rm YMS}(1,4;2,3|\vec{\pmb{\sigma}}_4)$ is determined by the property that YMS amplitudes with only two external scalars have vanishing soft behavior at the $\tau^{-1}$ order, as discussed in section \ref{subsec-introsoft}. Thus, the expectation is satisfied, and the consistency is ensured. Secondly, as discussed in section \ref{subsec-softbehavior}, each term in the full amplitude can be detected by considering $k_i\to\tau k_i$ for at least one leg $i$ with $i\in\{1,2,3,4\}$. Imposing cyclic invariance makes the resulted object involve all of ${\cal P}^i_{\rm YM^{+2}}(\vec{\pmb{\sigma}}_4)$ automatically, thus leading to the complete amplitude. Thirdly, the expansion formula in \eref{4p-amp} is independent of the ordering $\vec{\pmb\sigma}_4$. Suppose this characteristic is assumed at the beginning, then the cyclic symmetry among external legs, which fixes the amplitude completely, can be enlarged to the invariance under all permutations of external legs, as can be seen in the expansion \eref{4p-amp}. However, we did not use this assumption and restricted ourselves to the cyclic symmetry. It means in the story of this paper, the independence on $\vec{\pmb\sigma}_4$ emerges, rather than assumed. From the CHY point of view, the independence on $\vec{\pmb\sigma}_4$ implies the corresponding CHY integrand factorizes as ${\cal I}_{\rm CHY}={\cal I}_L{\cal I}_R$, where ${\cal I}_R$ is the Parke-Taylor factor for the ordering $\vec{\pmb\sigma}_4$. Therefore, the emergence of independence on $\vec{\pmb\sigma}_4$ is equivalent to the emergence of the double copy structure. Finally, the resulted amplitude in \eref{4p-amp} is expressed as the expansion to YMS and BAS amplitudes which incorporate only trivalent interactions. Such expansion coincides with the statement at the beginning of the previous section \ref{subsec-softbehavior} that all vertices under consideration in this paper can be decomposed into cubic ones.

\subsection{General YM$^{+2}$ amplitudes}
\label{subsec-YM+2-np}

Repeating the same manipulation of section \ref{subsec-YM+2-4p} recursively, one can obtain
the expansion formula for the general ${\rm YM}^{+2}$ amplitude with arbitrary number of external legs
\bea
{\cal A}_{\rm YM^{+2}}(\vec{\pmb{\sigma}}_n)=\sum_{\vec{\pmb{\rho}}}\,{\rm tr}(F_{\vec{\pmb{\rho}}})\,
{\cal A}_{\rm YMS}(\vec{\pmb{\rho}};\pmb{g}_n\setminus\pmb\rho|\vec{\pmb{\sigma}}_n)\,,~~~\label{np-amp}
\eea
where $\pmb{g}_n$ stands for the unordered
set of $n$ external gluons. Again, the effective part of summation is for inequivalent $\vec{\pmb{\rho}}$ which contain at least two elements.
At the end of this subsection, we will show the above expansion is coincide with that given in \cite{Bonnefoy:2023imz}, which is found
via the so called
covariant color-kinematic (CK) duality method based on the classical equation of motion.
The expansion for $4$-point amplitudes in \eref{4p-amp} manifestly satisfies the
general formula \eref{np-amp}. Now we show that \eref{np-amp} is correct for the $(m+1)$-point case if the $m$-point case is satisfied.
Some details and explanations will be omitted since they are paralleled to treatments for the $4$-point case in the previous subsection.

The subleading soft theorem for YM amplitudes indicates that
\bea
{\cal A}^{(1)_i}_{\rm YM^{+2}}(\vec{\pmb{\sigma}}_{m+1})&=&S^{(1)_i}_g\,{\cal A}_{\rm YM^{+2}}(\vec{\pmb{\sigma}}_{m+1}\setminus i)\,.
\eea
One can figure out the subleading contribution ${\cal A}^{(1)_i}_{\rm YM^{+2}}(\vec{\pmb{\sigma}}_{m+1})$ as
\bea
{\cal A}^{(1)_i}_{\rm YM^{+2}}(\vec{\pmb{\sigma}}_{m+1})
&=&S^{(1)_i}_g\,\Big(\sum_{\vec{\pmb{\rho}}}\,{\rm tr}(F_{\vec{\pmb{\rho}}})\,
{\cal A}_{\rm YMS}(\vec{\pmb{\rho}};\pmb{g}_{m+1}\setminus \{\pmb\rho\cup i\}|\vec{\pmb{\sigma}}_{m+1}\setminus i)\Big)\nn
&=&\sum_{\vec{\pmb{\rho}}}\,{\rm tr}(F_{\vec{\pmb{\rho}}})\,\Big(S^{(1)_i}_g\,
{\cal A}_{\rm YMS}(\vec{\pmb{\rho}};\pmb{g}_{m+1}\setminus \{\pmb\rho\cup i\}|\vec{\pmb{\sigma}}_{m+1}\setminus i)\Big)\nn
& &+\sum_{\vec{\pmb{\rho}}}\,\Big(S^{(1)_i}_g\,{\rm tr}(F_{\vec{\pmb{\rho}}})\Big)\,
{\cal A}_{\rm YMS}(\vec{\pmb{\rho}};\pmb{g}_{m+1}\setminus \{\pmb\rho\cup i\}|\vec{\pmb{\sigma}}_{m+1}\setminus i)\nn
&=&\sum_{\vec{\pmb{\rho}}}\,{\rm tr}(F_{\vec{\pmb{\rho}}})\,
{\cal A}^{(1)_i}_{\rm YMS}(\vec{\pmb{\rho}};\pmb{g}_{m+1}\setminus \pmb\rho|\vec{\pmb{\sigma}}_{m+1})\nn
& &+\tau\,\sum_{\vec{\pmb{\rho}}}\,{\rm tr}(F_{\rho_1,\vec{\pmb\gamma}\shuffle i})\,
{\cal A}^{(0)_i}_{\rm YMS}(\rho_1,\vec{\pmb\gamma}\shuffle i;\pmb{g}_{m+1}\setminus \{\pmb\rho\cup i\}|\vec{\pmb{\sigma}}_{m+1})\,,
\eea
which indicates
\bea
{\cal P}^i_{\rm YM^{+2}}(\vec{\pmb{\sigma}}_{m+1})
&=&\sum_{\vec{\pmb{\rho}}}\,{\rm tr}(F_{\vec{\pmb{\rho}}})\,
{\cal A}_{\rm YMS}(\vec{\pmb{\rho}};\pmb{g}_{m+1}\setminus\pmb\rho|\vec{\pmb{\sigma}}_{m+1})\nn
& &+\sum_{\vec{\pmb{\rho}}}\,{\rm tr}(F_{\rho_1,\vec{\pmb\gamma}\shuffle i})\,
{\cal A}_{\rm YMS}(\rho_1,\vec{\pmb\gamma}\shuffle i;\pmb{g}_{m+1}\setminus \{\pmb\rho\cup i\}|\vec{\pmb{\sigma}}_{m+1})\,.
\eea
Here $\vec{\pmb{\rho}}$ is an ordered set which contains at least two elements in $\pmb{g}_{m+1}\setminus i$,
and is labeled as $\vec{\pmb{\rho}}=\{\rho_1,\vec{\pmb\gamma}\}$, where $\rho_1$ is the first element in the ordering.
The shuffle $\vec{\pmb\a}\shuffle\vec{\pmb\b}$ means summing over all permutations such that the relative order in each of the ordered sets
$\vec{\pmb\a}$ and $\vec{\pmb\b}$ is kept. For instance, suppose $\vec{\pmb\a}=\{1,2\}$, $\vec{\pmb\b}=\{3,4\}$, then
$\vec{\pmb\a}\shuffle\vec{\pmb\b}$ is understood as
\bea
{\cal A}(\vec{\pmb\a}\shuffle\vec{\pmb\b})={\cal A}(1,2,3,4)+{\cal A}(1,3,2,4)+{\cal A}(1,3,4,2)\,.
\eea
It is straightforward to rewrite
${\cal P}^i_{\rm YM^{+2}}(\vec{\pmb{\sigma}}_{m+1})$ as
\bea
{\cal P}^i_{\rm YM^{+2}}(\vec{\pmb{\sigma}}_{m+1})&=&\sum_{\substack{\vec{\pmb\rho}',|\pmb{\rho}'|=2\\i\not{\in}\pmb{\rho}}}\,
{\rm tr}(F_{\vec{\pmb{\rho}}'})\,{\cal A}_{\rm YMS}(\vec{\pmb{\rho}};\pmb{g}_{m+1}\setminus\pmb\rho'|\vec{\pmb{\sigma}}_{m+1})\nn
& &+\sum_{\vec{\pmb\rho}',|\pmb{\rho}'|\geq3}\,{\rm tr}(F_{\vec{\pmb{\rho}}'})\,
{\cal A}_{\rm YMS}(\vec{\pmb{\rho}}';\pmb{g}_{m+1}\setminus \pmb\rho'|\vec{\pmb{\sigma}}_{m+1})\,,~~\label{mp-part}
\eea
where $|\pmb{\rho}'|$ denotes the number of elements in $\pmb{\rho}'$.
Requiring the invariance under cyclic permutation turns \eref{mp-part}
to
\bea
{\cal A}_{\rm YM^{+2}}(\vec{\pmb{\sigma}}_{m+1})&=&\sum_{\vec{\pmb{\rho}}'}\,
{\rm tr}(F_{\vec{\pmb{\rho}}'})\,{\cal A}_{\rm YMS}(\vec{\pmb{\rho}}';\pmb{g}_{m+1}\setminus\pmb\rho'|\vec{\pmb{\sigma}}_{m+1})\,,
\eea
which satisfies the general formula \eref{np-amp}.

The general expansion formula \eref{np-amp} manifests the permutation
symmetry among external gluons and the gauge invariance for each polarization, but is highly redundant. First, each pair of terms
\bea
{\rm tr}(F_{\vec{\pmb\rho}})\,{\cal A}_{\rm YMS}(\vec{\pmb\rho};\pmb{g}_{n}\setminus\pmb\rho|\vec{\pmb\sigma}_n)~~{\rm and}~~
{\rm tr}(F_{\vec{\pmb\rho}^T})\,{\cal A}_{\rm YMS}(\vec{\pmb\rho}^T;\pmb{g}_{n}\setminus\pmb\rho|\vec{\pmb\sigma}_n)~~\label{rela-equi}
\eea
are equivalent to each other, where the ordered set $\vec{\pmb\rho}^T$ is the reversing of $\vec{\pmb\rho}$.
This is because of the ordered reversed relations
\bea
{\rm tr}(F_{\vec{\pmb\rho}^T})=(-)^{|\vec{\pmb\rho}|}\,{\rm tr}(F_{\vec{\pmb\rho}})\,,
~~{\rm and}~~{\cal A}_{\rm YMS}(\vec{\pmb\rho}^T;\pmb{g}_{n}\setminus\pmb\rho|\vec{\pmb\sigma}_n)
=(-)^{|\vec{\pmb\rho}|}\,{\cal A}_{\rm YMS}(\vec{\pmb\rho};\pmb{g}_{n}\setminus\pmb\rho|\vec{\pmb\sigma}_n)\,,
\eea
where the first relation is based on the antisymmetry of $f_i^{\mu\nu}$, and the second one is a special case of the general relation
\bea
{\cal A}(1,\cdots,k)=(-)^k\,{\cal A}(k,\cdots,1)\,,~~\label{order-rever}
\eea
which is valid for any ordered amplitude. Secondly, various terms in \eref{np-amp} cancel each other. An example will be seen in sections \ref{subsec-YMS+2-4p} and \ref{subsec-differential} ahead.
As another example of cancelation, we now show that \eref{np-amp} can be reduced to
\bea
{\cal A}_{\rm YM^{+2}}(\vec{\pmb\sigma}_n)=\sum_{\vec{\pmb{\rho}},f\not\in\pmb\rho}\,
{\rm tr}(F_{\vec{\pmb{\rho}}})\,{\cal A}_{\rm YMS}(\vec{\pmb{\rho}};\pmb{g}_n\setminus\pmb\rho|\vec{\pmb{\sigma}}_n)\,,~~\label{np-equi}
\eea
where all terms with $f\in\pmb\rho$ are eliminated, for arbitrary $f\in\{1,\cdots,n\}$. Let us collect all terms with $f$ and another leg $i$
are fixed at two ends of $\vec{\pmb\rho}$, to get
\bea
G_{fi}&=&{\rm tr}(f_i\cdot f_f)\,{\cal A}_{\rm YMS}(f,i;\pmb{g}_n\setminus\{f,i\}|\vec{\pmb\sigma}_n)\nn
& &+2\,\sum_{\vec{\pmb\a}}\,{\rm tr}(f_i\cdot F_{\vec{\pmb\a}}\cdot f_f)\,
{\cal A}_{\rm YMS}(f,\vec{\pmb\a},i;\pmb{g}_n\setminus\{\{f,i\}\cup\pmb\a\}|\vec{\pmb\sigma}_n)\,.
\eea
Using the expansion of YM amplitudes with gauge invariant coefficients found in \cite{Hu:2023lso},
we get $G_{fi}=-(k_f\cdot k_i){\cal A}_{\rm YM}(\vec{\pmb\sigma}_n)$. Thus, summing over all $G_{fi}$ gives
\bea
\sum_{i\neq f}\,G_{fi}=-\sum_{i\neq f}\,(k_f\cdot k_i)\,{\cal A}_{\rm YM}(\vec{\pmb\sigma}_n)=0\,.~~\label{vanish}
\eea
Consequently, one can remove $\sum_{i\neq f}\,G_{fi}$ in \eref{np-amp} to obtain \eref{np-equi}. The simplified new version in\eref{np-equi}
is the expansion found in \cite{Bonnefoy:2023imz},
it breaks the manifest permutation invariance due to the fiducial gluon $f$, but contains less terms.

\section{Tree amplitudes of YMS$^{+2}$}
\label{sec-YMS+2}

In this section, we continue constructing the YMS$^{+2}$ amplitudes, which describe the scattering of gluons and BAS scalars with the single insertion of the operator $F^3$. In section \ref{subsec-YMS+2-4p}, we derive the $4$-point amplitudes of such a theory from the $4$-point ${\rm YM}^{+2}$ amplitudes constructed in the previous section using the dimensional reduction technique. Then, in section \ref{subsec-YMS+2-n+2p}, we construct YMS$^{+2}$ amplitudes with $2$ external gluons and an arbitrary number of external scalars by inverting the leading soft theorem for BAS scalars. Subsequently, we build general YMS$^{+2}$ amplitudes by inverting the subleading soft theorem for gluons in section \ref{subsec-YMS+2-general}. Finally, in section \ref{subsec-differential}, we verify that the differential operators, which transmute YM amplitudes to YMS ones, also connect ${\rm YM}^{+2}$ and ${\rm YMS}^{+2}$ amplitudes in exactly the same manner.

\subsection{$4$-point amplitudes}
\label{subsec-YMS+2-4p}

As discussed in section.\ref{subsec-softbehavior},
start from the $3$-point YM amplitude, one can generate the $3$-point
YMS amplitude with two external scalars through the dimensional reduction. For the ${\rm YM}^{+2}$ case,
we do not have such luxury, since the above operation annihilates the $3$-point ${\rm YM}^{+2}$ amplitude.
In other words, the consistent $3$-point ${\rm YMS}^{+2}$ amplitude does not exist.

However, one can use the dimensional reduction method to construct the well defined
$4$-point ${\rm YMS}^{+2}$ amplitudes with two external scalars and two external gluons.
To do this, we denote $4$ external legs of ${\cal A}_{\rm YM^{+2}}(\vec{\pmb{\sigma}}_4)$
as $1$, $2$, $a$, $b$, let $\epsilon_1$ and $\epsilon_2$ to lie in the extra dimension,
and assume $\epsilon_1\cdot\epsilon_2=1$ for simplicity.
The un-vanishing terms are those contain $\epsilon_1\cdot\epsilon_2$.
Use the expansion of $4$-point ${\rm YM}^{+2}$ amplitude in \eref{4p-amp}, we find the survived terms can be collected as
\bea
P_1&=&{\rm tr}(f_2\cdot f_1)\,{\cal A}_{\rm YMS}(1,2;\{a,b\}|\vec{\pmb{\sigma}}_4)\nn
& &+{\rm tr}(f_2\cdot f_a\cdot f_1)\,{\cal A}_{\rm YMS}(1,a,2;b|\vec{\pmb{\sigma}}_4)
+{\rm tr}(f_2\cdot f_b\cdot f_1)\,{\cal A}_{\rm YMS}(1,b,2;a|\vec{\pmb{\sigma}}_4)\nn
& &+{\rm tr}(f_a\cdot f_2\cdot f_1)\,{\cal A}_{\rm YMS}(1,2,a;b|\vec{\pmb{\sigma}}_4)
+{\rm tr}(f_b\cdot f_2\cdot f_1)\,{\cal A}_{\rm YMS}(1,2,b;a|\vec{\pmb{\sigma}}_4)\nn
& &+{\rm tr}(f_2\cdot f_b\cdot f_a\cdot f_1)\,{\cal A}_{\rm BAS}(1,a,b,2|\vec{\pmb{\sigma}}_4)
+{\rm tr}(f_2\cdot f_a\cdot f_b\cdot f_1)\,{\cal A}_{\rm BAS}(1,b,a,2|\vec{\pmb{\sigma}}_4)\nn
& &+{\rm tr}(f_b\cdot f_a\cdot f_2\cdot f_1)\,{\cal A}_{\rm BAS}(1,2,a,b|\vec{\pmb{\sigma}}_4)
+{\rm tr}(f_a\cdot f_b\cdot f_2\cdot f_1)\,{\cal A}_{\rm BAS}(1,2,b,a|\vec{\pmb{\sigma}}_4)\,,
\eea
and
\bea
P_2={\rm tr}(f_b\cdot f_a)\,{\cal A}_{\rm YMS}(a,b;\{1,2\}|\vec{\pmb{\sigma}}_4)\,.~~\label{P-2}
\eea
The dimensional reduction transmutes $P_1$ to
\bea
\W P_1&=&-2k_2\cdot k_1\,{\cal A}_{\rm YMS}(1,2;\{a,b\}|\vec{\pmb{\sigma}}_4)\nn
& &-2k_2\cdot f_a\cdot k_1\,{\cal A}_{\rm YMS}(1,a,2;b|\vec{\pmb{\sigma}}_4)
-2k_2\cdot f_b\cdot k_1\,{\cal A}_{\rm YMS}(1,b,2;a|\vec{\pmb{\sigma}}_4)\nn
& &-2k_2\cdot f_b\cdot f_a\cdot k_1\,{\cal A}_{\rm BAS}(1,a,b,2|\vec{\pmb{\sigma}}_4)
-2k_2\cdot f_a\cdot f_b\cdot k_1\,{\cal A}_{\rm BAS}(1,b,a,2|\vec{\pmb{\sigma}}_4)\,.
\eea
Using the expansion of YM amplitudes in \cite{Fu:2017uzt,Feng:2019tvb,Hu:2023lso}, we see that the above $\W P_1$ can be generated from
${\cal A}_{\rm YM}(\vec{\pmb{\sigma}}_4)$ via the replacement $\epsilon_1\to k_1$,
$\epsilon_2\to k_2$, up to an overall $-2$, thus the gauge invariance indicates $\W P_1=0$. Using the expansion
\bea
{\cal A}_{\rm YMS}(a,b;\{1,2\}|\vec{\pmb{\sigma}}_4)&=&(\epsilon_2\cdot k_a)\,{\cal A}_{\rm YMS}(a,2,b;1|\vec{\pmb{\sigma}}_4)\nn
& &+(\epsilon_2\cdot f_1\cdot k_a)\,{\cal A}_{\rm BAS}(a,1,2,b|\vec{\pmb{\sigma}}_4)\,,
\eea
we find that the dimensional reduction transmutes $P_2$ to
\bea
\W P_2=-{\rm tr}(f_b\cdot f_a)\,(k_a\cdot k_1)\,{\cal A}_{\rm BAS}(a,1,2,b|\vec{\pmb{\sigma}}_4)
=-{\rm tr}(f_b\cdot f_a)\,(k_a\cdot k_1)\,{\cal A}_{\rm BAS}(1,a,b,2|\vec{\pmb{\sigma}}_4)\,,~~\label{P2}
\eea
where the second equality uses the cyclic invariance of ordering, as well as the ordered reversed relation in \eref{order-rever}.
Using $\W P_1=0$ and $\W P_2$ in \eref{P2}, we arrive at
\bea
{\cal A}_{\rm YMS^{+2}}(1,2;\{a,b\}|\vec{\pmb{\sigma}}_4)=\W P_1+\W P_2=-{\rm tr}(f_b\cdot f_a)\,(k_a\cdot k_1)\,
{\cal A}_{\rm BAS}(1,a,b,2|\vec{\pmb{\sigma}}_4)\,.~~\label{4p-YMS+2}
\eea
\begin{figure}
  \centering
  \includegraphics[width=6cm]{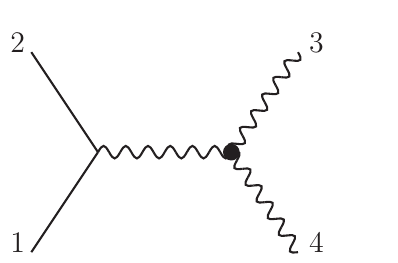} \\
  \caption{The only factorization channel for $4$-point ${\rm YMS}^{+2}$ amplitude with two external scalars and two external gluons.}\label{DR2}
\end{figure}

The result \eref{4p-YMS+2} carries only one pole $1/s_{12}$, as the pole $1/s_{1a}$ is canceled by $k_a\cdot k_1$ in the coefficient. This feature can be understood as follows: since the $3$-point ${\rm YM}^{+2}$ amplitude does not exist and two scalars can couple to a gluon, as shown in Figure \ref{DR1}, the amplitude ${\cal A}_{\rm YMS^{+2}}(1,2;{a,b}|\vec{\pmb{\sigma}}_4)$ has only one allowed factorization channel corresponding to $1/s_{12}$, as illustrated in Figure \ref{DR2}.

Since the amplitude \eref{4p-YMS+2} arises from the $P_2$ part in \eref{P-2}, it can be represented as
\bea
{\cal A}_{\rm YMS^{+2}}(1,2;\{a,b\}|\vec{\pmb{\sigma}}_4)&=&\partial_{\epsilon_1\cdot\epsilon_2}\,
\Big({\rm tr}(f_b\cdot f_a)\,{\cal A}_{\rm YMS}(a,b;\{1,2\}|\vec{\pmb\sigma}_4)\Big)\nn
&=&{\rm tr}(f_b\cdot f_a)\,\Big(\partial_{\epsilon_1\cdot\epsilon_2}\,\partial_{\epsilon_a\cdot\epsilon_b}\,
{\cal A}_{\rm YM}(\vec{\pmb\sigma}_4)\Big)\,,~~\label{reform-4p-YMS+2}
\eea
where the differential operator $\partial_{\epsilon_i\cdot\epsilon_j}$ is equivalent to the dimensional
reduction which turns gluons $i$ and $j$ to scalars, due to the linearity on each polarization.
This formula will be useful in the discussion of section \ref{subsec-YM+4-4p}.

\subsection{$(n+2)$-point amplitudes with $n$ external scalars}
\label{subsec-YMS+2-n+2p}

Start from the $4$-point amplitude \eref{4p-YMS+2}, one can invert the soft theorem for BAS scalar
in \eref{soft-s1} and \eref{soft-fac-s1}, to construct
amplitudes with more external scalars. In the resulted higher-point amplitudes, external scalars carry
two orderings, one is among all external legs, while another one is among only external scalars without
including any gluon, thus we interpret such amplitudes as BAS scalars couple to gluons.

Let us construct the $5$-point amplitude ${\cal A}_{\rm YMS^{+2}}(1,2,3;\{a,b\}|\vec{\pmb\sigma}_5)$ at first.
Consider $k_2\to\tau k_2$, the soft theorem
forces
\bea
{\cal A}^{(0)_2}_{\rm YMS^{+2}}(1,2,3;\{a,b\}|\vec{\pmb\sigma}_5)&=&S^{(0)_2}_s\,
{\cal A}_{\rm YMS^{+2}}(1,3;\{a,b\}|\vec{\pmb\sigma}_5\setminus 2)\nn
&=&-{1\over\tau}\,\Big({\delta_{12}\over s_{12}}+{\delta_{23}\over s_{23}}\Big)\,
\Big({\rm tr}(f_b\cdot f_a)\,(k_a\cdot k_1)\,{\cal A}_{\rm BAS}(1,a,b,3|\vec{\pmb{\sigma}}_5\setminus 2)\Big)\nn
&=&-{\rm tr}(f_b\cdot f_a)\,(k_a\cdot k_1)\,{\cal A}^{(0)_2}_{\rm BAS}(1,\{a,b\}\shuffle 2,3|\vec{\pmb{\sigma}}_5)\,.~~~\label{5p-soft}
\eea
In \eref{5p-soft}, the second equality uses the expansion in \eref{4p-YMS+2} and the soft factor in \eref{soft-fac-s1}.
To get the third one, we have used the property in \eref{tech}, which leads to
\bea
& &{\cal A}^{(0)_2}_{\rm BAS}(1,2,a,b,3|\vec{\pmb{\sigma}}_5)+{\cal A}^{(0)_2}_{\rm BAS}(1,a,2,b,3|\vec{\pmb{\sigma}}_5)
+{\cal A}^{(0)_2}_{\rm BAS}(1,a,b,2,3|\vec{\pmb{\sigma}}_5)\nn
&=&{1\over\tau}\,\Big({\delta_{12}\over s_{12}}+{\delta_{2a}\over s_{2a}}+{\delta_{a2}\over s_{a2}}
+{\delta_{2b}\over s_{2b}}+{\delta_{b2}\over s_{b2}}+{\delta_{23}\over s_{23}}\Big)\,{\cal A}_{\rm BAS}(1,a,b,3|\vec{\pmb{\sigma}}_5\setminus 2)\nn
&=&{1\over\tau}\,\Big({\delta_{12}\over s_{12}}+{\delta_{23}\over s_{23}}\Big)\,
{\cal A}_{\rm BAS}(1,a,b,3|\vec{\pmb{\sigma}}_5\setminus 2)\,.
\eea
The soft behavior in \eref{5p-soft} indicates that ${\cal A}_{\rm YMS^{+2}}(1,2,3;\{a,b\}|\vec{\pmb\sigma}_5)$ can be expanded to BAS amplitudes as
\bea
{\cal A}_{\rm YMS^{+2}}(1,2,3;\{a,b\}|\vec{\pmb\sigma}_5)&=&C(\shuffle)\,
{\cal A}_{\rm BAS}(1,\{a,b\}\shuffle 2,3|\vec{\pmb{\sigma}}_5)\,,~~\label{expan-5p}
\eea
where the coefficients $C(\shuffle)$ satisfy
\bea
C(\shuffle)\Big|_{k_2\to\tau k_2,\tau\to0}=-{\rm tr}(f_b\cdot f_a)\,(k_a\cdot k_1)\,.~~\label{C_1}
\eea

To fix $C(\shuffle)$ completely, we consider the soft behavior for $k_1\to\tau k_1$ to get
\bea
{\cal A}^{(0)_1}_{\rm YMS^{+2}}(1,2,3;\{a,b\}|\vec{\pmb\sigma}_5)&=&S^{(0)_1}_s\,
{\cal A}_{\rm YMS^{+2}}(2,3;\{a,b\}|\vec{\pmb\sigma}_5\setminus 1)\nn
&=&-{1\over\tau}\,\Big({\delta_{31}\over s_{31}}+{\delta_{12}\over s_{12}}\Big)\,\Big({\rm tr}(f_b\cdot f_a)\,(k_a\cdot k_2)\,
{\cal A}_{\rm BAS}(2,a,b,3|\vec{\pmb{\sigma}}_5\setminus 1)\Big)\nn
&=&-{\rm tr}(f_b\cdot f_a)\,(k_a\cdot k_2)\,{\cal A}^{(0)_2}_{\rm BAS}(1,2,a,b,3|\vec{\pmb{\sigma}}_5)\,,~~~\label{5p-soft-1}
\eea
which means
\bea
& &C(2,a,b)\Big|_{k_1\to\tau k_1,\tau\to0}=-{\rm tr}(f_b\cdot f_a)\,(k_a\cdot k_2)\,,\nn
& &C(a,2,b)\Big|_{k_1\to\tau k_1,\tau\to0}
=C(a,b,2)\Big|_{k_1\to\tau k_1,\tau\to0}=0\,.~~~\label{C-2}
\eea
Comparing \eref{C-2} with \eref{C_1}, we find
\bea
C(\shuffle)=-{\rm tr}(f_b\cdot f_a)\,(k_a\cdot Y_a)\,,~~\label{define-C}
\eea
where $Y_a$ is defined as the summation of external momenta carried by external scalars at the l.h.s of $a$ in the ordering.
As can be verified directly, with coefficients $C(\shuffle)$ defined in \eref{define-C}, the expansion formula \eref{expan-5p} also gives
the correct soft behavior for $k_3\to\tau k_3$. Thus, we conclude that the $5$-point ${\rm YM}^{+2}$ amplitude is given as
\bea
{\cal A}_{\rm YMS^{+2}}(1,2,3;\{a,b\}|\vec{\pmb\sigma}_5)&=&-{\rm tr}(f_b\cdot f_a)\,(k_a\cdot Y_a)\,
{\cal A}_{\rm BAS}(1,\{a,b\}\shuffle 2,3|\vec{\pmb{\sigma}}_5)\,.~~\label{5p-amp}
\eea

Repeating the above construction recursively, one can find the $(n+2)$-point ${\rm YMS}^{+2}$ amplitude
with $n$ external scalars and $2$ external gluons to be
\bea
{\cal A}_{\rm YMS^{+2}}(1,\cdots,n;\{a,b\}|\vec{\pmb{\sigma}}_{n+2})=-{\rm tr}(f_b\cdot f_a)\,(k_a\cdot Y_a)\,{\cal A}_{\rm BAS}
(1,\{2,\cdots,n-1\}\shuffle\{a,b\},n|\vec{\pmb{\sigma}}_{n+2})\,.~~\label{n+2p-amp}
\eea
It seems that this formula breaks the permutation symmetry among external gluons $a$ and $b$. Such symmetry is protected by the relation
\bea
& &(k_a\cdot X_a)\,{\cal A}_{\rm BAS}
(1,\{2,\cdots,n-1\}\shuffle\{a,b\},n|\vec{\pmb{\sigma}}_{n+2})\nn
&=&(k_b\cdot X_b)\,{\cal A}_{\rm BAS}
(1,\{2,\cdots,n-1\}\shuffle\{b,a\},n|\vec{\pmb{\sigma}}_{n+2})\,,~~\label{rela}
\eea
where $X_a$ (or $X_b$) is defined as the summation of external momenta carried by legs at the l.h.s of $a$ (or $b$) in the ordering,
without distinguishing scalar or gluon. For the above case, $X_a$ (or $X_b$) is equivalent to $Y_a$ (or $Y_b$).
The relation \eref{rela} can be proved as follows. The fundamental BCJ relation yields
\bea
& &(k_a\cdot X_a)\,{\cal A}_{\rm BAS}
(1,\{2,\cdots,n-1\}\shuffle\{a,b\},n|\vec{\pmb{\sigma}}_{n+2})\nn
&=&-(k_a\cdot X_a)\,{\cal A}_{\rm BAS}
(1,\{2,\cdots,n-1\}\shuffle\{b,a\},n|\vec{\pmb{\sigma}}_{n+2})\,,
\eea
then \eref{rela} is governed by the general BCJ relation
\bea
\Big(k_a\cdot X_a+k_b\cdot X_b\Big)\,{\cal A}_{\rm BAS}
(1,\{2,\cdots,n-1\}\shuffle\{b,a\},n|\vec{\pmb{\sigma}}_{n+2})=0\,.
\eea
%

\subsection{General YMS$^{+2}$ amplitudes}
\label{subsec-YMS+2-general}

Now we can continue to construct the more general ${\rm YMS}^{+2}$ amplitude
${\cal A}_{\rm YMS^{+2}}(1,\cdots,n;\pmb{g}_m|\vec{\pmb{\sigma}}_{n+m})$
with $n$ external scalars and $m$ external gluons. The mass dimension implies that
${\cal A}_{\rm YMS^{+2}}(1,\cdots,n;\pmb{g}_m|\vec{\pmb{\sigma}}_{n+m})$
contains only one ${\rm YM}^{+2}$ vertex, thus it can be generated by inserting external gluons attached to YM vertices.
Similar to section \ref{sec-YM+2}, our main technique in this subsection is to realize such insertion by inverting the subleading soft theorem for YM amplitudes.

To start, let us consider the amplitude ${\cal A}_{\rm YMS^{+2}}(1,\cdots,n;\{a,b,c\}|\vec{\pmb{\sigma}}_{n+3})$
with $n$ external scalars and $3$ external gluons $a$, $b$ and $c$. We can take $k_c\to\tau k_c$, then the subleading soft theorem forces
\bea
&&{\cal A}^{(1)_c}_{\rm YMS^{+2}}(1,\cdots,n;\{a,b,c\}|\vec{\pmb{\sigma}}_{n+3})=S^{(1)_c}_g\,
{\cal A}_{\rm YMS^{+2}}(1,\cdots,n;\{a,b\}|\vec{\pmb{\sigma}}_{n+3}\setminus c)\nn
&&~~~~~~~~~~~~~~~=-{\rm tr}(f_b\cdot f_a)\,(k_a\cdot Y_a)\,\Big(S^{(1)_c}_g\,
{\cal A}_{\rm BAS}(1,\{2,\cdots,n-1\}\shuffle\{a,b\},n|\vec{\pmb{\sigma}}_{n+3}\setminus c)\Big)\nn
&&~~~~~~~~~~~~~~~~~~-{\rm tr}(f_b\cdot f_a)\,\Big(S^{(1)_c}_g\,(k_a\cdot Y_a)\Big)\,
{\cal A}_{\rm BAS}(1,\{2,\cdots,n-1\}\shuffle\{a,b\},n|\vec{\pmb{\sigma}}_{n+3}\setminus c)\nn
&&~~~~~~~~~~~~~~~~~~-\Big(S^{(1)_c}_g\,{\rm tr}(f_b\cdot f_a)\Big)\,(k_a\cdot Y_a)\,
{\cal A}_{\rm BAS}(1,\{2,\cdots,n-1\}\shuffle\{a,b\},n|\vec{\pmb{\sigma}}_{n+3}\setminus c)\nn
&&~~~~~~~~~~~~~~~=B_1+B_2+B_3\,.~~\label{B1B2B3}
\eea
In the above, the $B_1$ part can be obtained by inverting the subleading soft theorem
\bea
B_1&=&-{\rm tr}(f_b\cdot f_a)\,(k_a\cdot Y_a)\,\Big(S^{(1)_c}_g\,
{\cal A}_{\rm BAS}(1,\{2,\cdots,n-1\}\shuffle\{a,b\},n|\vec{\pmb{\sigma}}_{n+3}\setminus c)\Big)\nn
&=&-{\rm tr}(f_b\cdot f_a)\,(k_a\cdot Y_a)\,{\cal A}^{(1)_c}_{\rm YMS}(1,\{2,\cdots,n-1\}\shuffle\{a,b\},n;c|\vec{\pmb{\sigma}}_{n+3})\,.
\eea
The $B_2$ part can be evaluated as
\bea
B_2&=&-{\rm tr}(f_b\cdot f_a)\,\Big(S^{(1)_c}_g\,(k_a\cdot Y_a)\Big)\,
{\cal A}_{\rm BAS}(1,\{2,\cdots,n-1\}\shuffle\{a,b\},n|\vec{\pmb{\sigma}}_{n+3}\setminus c)\nn
&=&-\tau\,{\rm tr}(f_b\cdot f_a)\,(k_a\cdot f_c\cdot Y_c)\,
{\cal A}^{(0)_c}_{\rm BAS}(1,\{2,\cdots,n-1\}\shuffle\{c,a,b\},n|\vec{\pmb{\sigma}}_{n+3})\,,~~\label{B2}
\eea
where the relation \eref{iden-1} and the soft theorem \eref{soft-s1} \eqref{soft-fac-s1} have been used. The $B_3$ part can be evaluated as
\bea
B_3&=&-\Big(S^{(1)_c}_g\,{\rm tr}(f_b\cdot f_a)\Big)\,(k_a\cdot Y_a)\,
{\cal A}_{\rm BAS}(1,\{2,\cdots,n-1\}\shuffle\{a,b\},n|\vec{\pmb{\sigma}}_{n+3}\setminus c)\nn
&=&-\tau\,{\rm tr}(f_b\cdot f_c\cdot f_a)\,(k_a\cdot Y_a)\,
{\cal A}^{(0)_c}_{\rm BAS}(1,\{2,\cdots,n-1\}\shuffle\{a,c,b\},n|\vec{\pmb{\sigma}}_{n+3})\nn
& &+\tau\,{\rm tr}(f_c\cdot f_b\cdot f_a)\,(k_a\cdot Y_a)\,
{\cal A}^{(0)_c}_{\rm BAS}(1,\{2,\cdots,n-1\}\shuffle\{a,c,b\},n|\vec{\pmb{\sigma}}_{n+3})\,,~~\label{B3}
\eea
via the similar technic.
The initial formula of the last line in \eref{B3} is
\bea
-{\rm tr}(f_c\cdot f_b\cdot f_a)\,(k_a\cdot Y_a)\,\Big({\delta_{bc}\over s_{bc}}
+{\delta_{ca}\over s_{ca}}\Big)\,{\cal A}_{\rm BAS}(1,\{2,\cdots,n-1\}\shuffle\{a,b\},n|\vec{\pmb{\sigma}}_{n+3}\setminus c)\,,
\eea
we have used $\delta_{bc}=-\delta_{cb}$ and $\delta_{ca}=-\delta_{ac}$, as well as the decomposition in \eref{tech},
to turn it to the formula in
the last line. The decomposition in \eref{tech} has also been used for obtaining \eref{B2} and another part in \eref{B3}.

Collecting $B_1$, $B_2$ and $B_3$ together, we get
\bea
& &{\cal A}^{(1)_c}_{\rm YMS^{+2}}(1,\cdots,n;a,b,c|\vec{\pmb{\sigma}}_{n+3})\nn
&=&-{\rm tr}(f_b\cdot f_a)\,(k_a\cdot Y_a)\,{\cal A}^{(1)_c}_{\rm YMS}(1,\{2,\cdots,n-1\}\shuffle\{a,b\},n;c|\vec{\pmb{\sigma}}_{n+3})\nn
& &-\tau\,{\rm tr}(f_b\cdot f_a)\,(k_a\cdot f_c\cdot Y_c)\,
{\cal A}^{(0)_c}_{\rm BAS}(1,\{2,\cdots,n-1\}\shuffle\{c,a,b\},n|\vec{\pmb{\sigma}}_{n+3})\nn
& &-\tau\,{\rm tr}(f_b\cdot f_c\cdot f_a)\,(k_a\cdot Y_a)\,
{\cal A}^{(0)_c}_{\rm BAS}(1,\{2,\cdots,n-1\}\shuffle\{a,c,b\},n|\vec{\pmb{\sigma}}_{n+3})\nn
& &+\tau\,{\rm tr}(f_c\cdot f_b\cdot f_a)\,(k_a\cdot Y_a)\,
{\cal A}^{(0)_c}_{\rm BAS}(1,\{2,\cdots,n-1\}\shuffle\{a,c,b\},n|\vec{\pmb{\sigma}}_{n+3})\,,
\eea
which indicates
\bea
& &{\cal P}^c_{\rm YMS^{+2}}(1,\cdots,n;\{a,b,c\}|\vec{\pmb{\sigma}}_{n+3})\nn
&=&-{\rm tr}(f_b\cdot f_a)\,(k_a\cdot Y_a)\,{\cal A}_{\rm YMS}(1,\{2,\cdots,n-1\}\shuffle\{a,b\},n;c|\vec{\pmb{\sigma}}_{n+3})\nn
& &-{\rm tr}(f_b\cdot f_a)\,(k_a\cdot f_c\cdot Y_c)\,{\cal A}_{\rm BAS}(1,\{2,\cdots,n-1\}\shuffle\{c,a,b\},n|\vec{\pmb{\sigma}}_{n+3})\nn
& &-{\rm tr}(f_b\cdot f_c\cdot f_a)\,(k_a\cdot Y_a)\,{\cal A}_{\rm BAS}(1,\{2,\cdots,n-1\}\shuffle\{a,c,b\},n|\vec{\pmb{\sigma}}_{n+3})\nn
& &+{\rm tr}(f_c\cdot f_b\cdot f_a)\,(k_a\cdot Y_a)\,{\cal A}_{\rm BAS}(1,\{2,\cdots,n-1\}\shuffle\{a,c,b\},n|\vec{\pmb{\sigma}}_{n+3})\,.
\eea

We have found ${\cal P}^c_{\rm YMS^{+2}}(1,\cdots,n;\{a,b,c\}|\vec{\pmb{\sigma}}_{n+3})$ which is a part of the full amplitude.
The complete amplitude can be achieved by imposing the permutation symmetry among gluons $a$, $b$ and $c$, similar as that discussed
in section \ref{subsec-YM+2-4p}. The permutation invariance yileds
\bea
& &{\cal A}_{\rm YMS^{+2}}(1,\cdots,n;a,b,c|\vec{\pmb{\sigma}}_{n+3})\nn
&=&-{\rm tr}(f_b\cdot f_a)\,(k_a\cdot Y_a)\,{\cal A}_{\rm YMS}(1,\{2,\cdots,n-1\}\shuffle\{a,b\},n;c|\vec{\pmb{\sigma}}_{n+3})\nn
& &-{\rm tr}(f_a\cdot f_c)\,(k_c\cdot Y_c)\,{\cal A}_{\rm YMS}(1,\{2,\cdots,n-1\}\shuffle\{c,a\},n;b|\vec{\pmb{\sigma}}_{n+3})\nn
& &-{\rm tr}(f_c\cdot f_b)\,(k_b\cdot Y_b)\,{\cal A}_{\rm YMS}(1,\{2,\cdots,n-1\}\shuffle\{b,c\},n;a|\vec{\pmb{\sigma}}_{n+3})\nn
& &-{\rm tr}(f_b\cdot f_a)\,(k_a\cdot f_c\cdot Y_c)\,{\cal A}_{\rm BAS}(1,\{2,\cdots,n-1\}\shuffle\{c,a,b\},n|\vec{\pmb{\sigma}}_{n+3})\nn
& &-{\rm tr}(f_a\cdot f_c)\,(k_c\cdot f_b\cdot Y_b)\,{\cal A}_{\rm BAS}(1,\{2,\cdots,n-1\}\shuffle\{b,c,a\},n|\vec{\pmb{\sigma}}_{n+3})\nn
& &-{\rm tr}(f_c\cdot f_b)\,(k_b\cdot f_a\cdot Y_a)\,{\cal A}_{\rm BAS}(1,\{2,\cdots,n-1\}\shuffle\{a,b,c\},n|\vec{\pmb{\sigma}}_{n+3})\nn
& &-{\rm tr}(f_b\cdot f_c\cdot f_a)\,(k_a\cdot Y_a)\,{\cal A}_{\rm BAS}(1,\{2,\cdots,n-1\}\shuffle\{a,c,b\},n|\vec{\pmb{\sigma}}_{n+3})\nn
& &+{\rm tr}(f_b\cdot f_c\cdot f_a)\,(k_c\cdot Y_c)\,{\cal A}_{\rm BAS}(1,\{2,\cdots,n-1\}\shuffle\{c,a,b\},n|\vec{\pmb{\sigma}}_{n+3})\nn
& &-{\rm tr}(f_c\cdot f_b\cdot f_a)\,(k_c\cdot Y_c)\,{\cal A}_{\rm BAS}(1,\{2,\cdots,n-1\}\shuffle\{c,a,b\},n|\vec{\pmb{\sigma}}_{n+3})\nn
& &+{\rm tr}(f_c\cdot f_b\cdot f_a)\,(k_a\cdot Y_a)\,{\cal A}_{\rm BAS}(1,\{2,\cdots,n-1\}\shuffle\{a,c,b\},n|\vec{\pmb{\sigma}}_{n+3})\,.
~~\label{n+3p-amp1}
\eea
The expansion in \eref{n+3p-amp1} does not manifest the permutation invariance.
Indeed, the permutation symmetry is ensured by various relations such as \eref{rela} and so on.
We postpone the proof of permutation symmetry of \eref{n+3p-amp1} to the next subsection.
For consistency, we need to verify that all undetectable terms in \eref{n+3p-amp1} vanish at the
$\tau^0$ order when considering $k_c\to\tau k_c$. As can be seen in \eref{n+3p-amp1},
most of undetectable terms are manifestly at the $\tau^1$ order, since their coefficients are bilinear in $k_c$, except the remaining terms
\bea
{\cal R}&=&-{\rm tr}(f_c\cdot f_b)\,(k_b\cdot Y_b)\,{\cal A}_{\rm YMS}(1,\{2,\cdots,n-1\}\shuffle\{b,c\},n;a|\vec{\pmb{\sigma}}_{n+3})\nn
& &-{\rm tr}(f_c\cdot f_b)\,(k_b\cdot f_a\cdot Y_a)\,{\cal A}_{\rm BAS}(1,\{2,\cdots,n-1\}\shuffle\{a,b,c\},n|\vec{\pmb{\sigma}}_{n+3})\,.
\eea
At the $\tau^0$ order, we have
\bea
{\cal R}^{(0)_c}&=&-\tau\,{\rm tr}(f_c\cdot f_b)\,S^{(0)_c}_s\,\Big((k_b\cdot Y_b)\,
{\cal A}_{\rm YMS}(1,\{2,\cdots,n-1\}\shuffle b,n;a|\vec{\pmb{\sigma}}_{n+3}\setminus c)\nn
& &~~~~~~~~~~~~~~~~~~~~~~~+(k_b\cdot f_a\cdot Y_a)\,
{\cal A}_{\rm BAS}(1,\{2,\cdots,n-1\}\shuffle\{a,b\},n|\vec{\pmb{\sigma}}_{n+3}\setminus c)\Big)\,.
\eea
Using the expansion
\bea
{\cal A}_{\rm YMS}(1,\cdots,n;\{a,b\})&=&(\epsilon_b\cdot Y_b)\,
{\cal A}_{\rm YMS}(1,\{2,\cdots,n-1\}\shuffle b,n;a|\vec{\pmb{\sigma}}_{n+3}\setminus c)\nn
& &+(\epsilon_b\cdot f_a\cdot Y_a)\,{\cal A}_{\rm BAS}(1,\{2,\cdots,n-1\}\shuffle\{a,b\},n|\vec{\pmb{\sigma}}_{n+3}\setminus c)\,,
\eea
we see the vanishing of ${\cal R}^{(0)_c}$, due to the gauge invariance for the polarization $\epsilon_b$.

The resulted amplitude \eref{n+3p-amp1} can be rearranged into the following more compact form
\bea
& &{\cal A}_{\rm YMS^{+2}}(1,\cdots,n;\{a,b,c\}|\vec{\pmb{\sigma}}_{n+3})\nn
&=&\sum_{\vec{\pmb{\a}}_1,\vec{\pmb{\a}}_2}\,\sum_{t\in\{\pmb\a_1\setminus f\}}\,
(-)^{|{\pmb\b}_1|+1}\,{\rm tr}(F_{\vec{\pmb{\a}}_1})\,(k_t\cdot F_{\vec{\pmb\a}_2}\cdot Y_{\vec{\pmb\a}_2})\nn
& &~~~~~~~~~~~~~~~~~~~~~~
{\cal A}_{\rm YMS}(1,\{2,\cdots,n-1\}\shuffle \{\vec{\pmb\a}_2,K^{\vec{\pmb\a}_1}_{t,f}\},n;\{a,b,c\}
\setminus\{\pmb\a_1\cup\pmb\a_2\}|\vec{\pmb{\sigma}}_{n+3})\,.~~\label{n+3p-compact}
\eea
This formula is chosen for latter convenience since it is valid for general ${\rm YMS}^{+2}$ amplitudes.
In \eref{n+3p-compact}, $\pmb\a_1$ and $\pmb\a_2$ are two subsets of $\{a,b,c\}$ without any overlap,
where $\pmb\a_1$ includes at least two elements and $\pmb\a_2$ is allowed to be empty
(when $\pmb\a_2=\emptyset$, $k_t\cdot F_{\vec{\pmb{\a}}_2}\cdot Y_{\vec{\pmb\a}_2}$ should be understood as
$k_t\cdot Y_t$). Giving orders to elements in $\pmb\a_1$ and $\pmb\a_2$, we get two corresponding ordered sets
$\vec{\pmb\a}_1$ and $\vec{\pmb\a}_2$. Assuming $\vec{\pmb{\a}}_1=\{\vec{\pmb\b}_1,t,\vec{\pmb\b}_2,f\}$, the ordered sets
$K^{\vec{\pmb\a}_1}_{t,f}$ are defined as $K^{\vec{\pmb\a}_1}_{t,f}=\{t,\vec{\pmb\b}_2\shuffle\vec{\pmb\b}_1^T,f\}$,
where $\vec{\pmb\b}_1^T$ is the reversing of $\vec{\pmb\b}_1$, and all possible shuffles should be summed.
As can be seen in \eref{n+3p-compact}, one should choose a fiducial leg $f$ in each $\pmb\a_1$, and sum over all
$t\in\{\pmb\a_1\setminus f\}$. Notice that the fiducial leg $f$ for each $\vec{\pmb\a}_1$ is chosen individually,
we do not require any two different $\vec{\pmb\a}_1$ to share the same fiducial leg.
All possible $\vec{\pmb\a}_1$ and $\vec{\pmb\a}_2$ should be summed over, and the fiducial leg $f$ avoids the
over counting of equivalent $\vec{\pmb\a}_1$ related by cyclic permutations. The result in \eref{n+3p-amp1}
explicitly satisfies the compact formula in \eref{n+3p-compact}.

Repeating the above manipulation, the general ${\rm YMS}^{+2}$ amplitude is found to be
\bea
& &{\cal A}_{\rm YMS^{+2}}(1,\cdots,n;\pmb{g}_m|\vec{\pmb{\sigma}}_{n+m})\nn
&=&\sum_{\vec{\pmb{\a}}_1,\vec{\pmb{\a}}_2}\,\sum_{t\in\{\pmb\a_1\setminus f\}}\,(-)^{|\vec{\pmb\b}_1|+1}\,
{\rm tr}(F_{\vec{\pmb{\a}}_1})\,(k_t\cdot F_{\vec{\pmb\a}_2}\cdot Y_{\vec{\pmb\a}_2})\nn
& &~~~~~~~~~~~~~~~~~~~~~~
{\cal A}_{\rm YMS}(1,\{2,\cdots,n-1\}\shuffle \{\vec{\pmb\a}_2,K^{\vec{\pmb\a}_1}_{t,f}\},n;
\pmb{g}_m\setminus\{\pmb\a_1\cup\pmb\a_2\}|\vec{\pmb{\sigma}}_{n+m})\,.~~\label{n+mp-amp}
\eea
Follow the manipulation paralleled to that from \eref{B1B2B3} to \eref{n+3p-amp1},
it is easy to show that if the general formula \eref{n+mp-amp}
is satisfied by the $(n+k)$-point case with $n$ external scalars and $k$ external gluons, then it is also satisfied by the $(n+k+1)$-point case.
In next subsection, we will re-derive the formula \eref{n+mp-amp} by using differential operators.

\subsection{Another derivation via transmutation operators}
\label{subsec-differential}

In section \ref{subsec-YMS+2-4p}, the $4$-point ${\rm YMS}^{+2}$ amplitude is generated from the $4$-point
${\rm YM}^{+2}$ one via the dimensional reduction. Obviously, the dimensional reduction is equivalent to applying the operator $\partial_{\epsilon_1\cdot\epsilon_2}$ to the $4$-point ${\rm YM}^{+2}$ amplitude,
as shown in \eref{reform-4p-YMS+2}. Thus, it is nature to ask if the general ${\rm YMS}^{+2}$ amplitude in \eref{n+mp-amp}
can be generated from the $(n+m)$-point ${\rm YM}^{+2}$ amplitude via more complicated operators.

On the other hand, tree level YMS and YM amplitudes can be linked as
\bea
{\cal A}_{\rm YMS}(1,\cdots,n;\pmb{g}_m|\vec{\pmb{\sigma}}_{n+m})={\cal T}[1,\cdots,n]\,{\cal A}_{\rm YM}(\vec{\pmb{\sigma}}_{n+m})\,,
~~\label{trans-operator}
\eea
where the combinatorial operator ${\cal T}[1,\cdots,n]$ is defined as
\bea
{\cal T}[1,\cdots,n]\equiv\partial_{\epsilon_1\cdot\epsilon_n}\,\prod_{i=2}^{n-1}\,
(\partial_{\epsilon_i\cdot k_{i-1}}-\partial_{\epsilon_i\cdot k_n})\,.~~\label{defin-oper}
\eea
The above operator and transmutation relation was first proposed by Cheung, Shen and Wen \cite{Cheung:2017ems},
then proved by employing CHY formula \cite{Zhou:2018wvn,
Bollmann:2018edb}. The proof in \cite{Zhou:2018wvn,
Bollmann:2018edb} is restricted to the ordinary YMS and YM case.
However, the initial construction for the operator ${\cal T}[1,\cdots,n]$ in \cite{Cheung:2017ems} is
based on the general consideration of gauge invariance and momentum conservation, without the aid of any particular theory.
Thus, it is nature to expect that the transmutation relation that the operator
${\cal T}[1,\cdots,n]$ turns gluons to scalars holds for a wider range of theories.

Now we show that the operator ${\cal T}[1,\cdots,n]$ do transmute the ${\rm YM}^{+2}$ amplitude to the ${\rm YMS}^{+2}$ amplitude as follows
\bea
{\cal A}_{\rm YMS^{+2}}(1,\cdots,n;\pmb{g}_m|\vec{\pmb{\sigma}}_{n+m})={\cal T}[1,\cdots,n]\,
{\cal A}_{\rm YM^{+2}}(\vec{\pmb{\sigma}}_{n+m})\,.~~\label{trans}
\eea
The operator $\partial_{\epsilon_1\cdot\epsilon_n}$ in ${\cal T}[1,\cdots,n]$ annihilates all terms do not contain $\epsilon_1\cdot\epsilon_n$,
thus, similar as in section \ref{subsec-YMS+2-4p}, the survived terms in ${\cal A}_{\rm YM^{+2}}(\vec{\pmb{\sigma}}_{n+m})$ can be collected as
\bea
P_1&=&{\rm tr}(f_n\cdot f_1)\,{\cal A}_{\rm YMS}(1,n;\pmb{g}_{n+m}\setminus\{1,n\}|\vec{\pmb{\sigma}}_{n+m})\nn
& &+\sum_{\vec{\pmb\a}}\,{\rm tr}(f_n\cdot F_{\vec{\pmb\a}}\cdot f_1)\,
{\cal A}_{\rm YMS}(1,\vec{\pmb\a},n;\pmb{g}_{n+m}\setminus\{\{1,n\}\cup\pmb\a\}|\vec{\pmb{\sigma}}_{n+m})\nn
& &+\sum_{\vec{\pmb\b}}\,{\rm tr}(f_1\cdot F_{\vec{\pmb\b}}\cdot f_n)\,
{\cal A}_{\rm YMS}(n,\vec{\pmb\b},1;\pmb{g}_{n+m}\setminus\{\{1,n\}\cup\pmb\b\}|\vec{\pmb{\sigma}}_{n+m})\nn
&=&{\rm tr}(f_n\cdot f_1)\,{\cal A}_{\rm YMS}(1,n;\pmb{g}_{n+m}\setminus\{1,n\}|\vec{\pmb{\sigma}}_{n+m})\nn
& &+2\,\sum_{\vec{\pmb\a}}\,{\rm tr}(f_n\cdot F_{\vec{\pmb\a}}\cdot f_1)\,
{\cal A}_{\rm YMS}(1,\vec{\pmb\a},n;\pmb{g}_{n+m}\setminus\{\{1,n\}\cup\pmb\a\}|\vec{\pmb{\sigma}}_{n+m})\,,~~\label{diff-P1}
\eea
and
\bea
P_2&=&\sum_{\vec{\pmb\gamma}}\,{\rm tr}(F_{\vec{\pmb\gamma}})\,
{\cal A}_{\rm YMS}(\vec{\pmb{\gamma}};\pmb{g}_{n+m}\setminus\pmb\gamma|\vec{\pmb{\sigma}}_{n+m})\,,~~\label{diff-P2}
\eea
where $\pmb\a$, $\pmb\b$ and $\pmb\gamma$ are un-empty subsets of $\pmb{g}_{n+m}\setminus \{1,n\}$, i.e.,
non of them includes the element $1$ or $n$.
The ordered sets $\vec{\pmb{\a}}$, $\vec{\pmb\b}$ and $\vec{\pmb\gamma}$ are obtained by giving orders to elements in
$\pmb\a$, $\pmb\b$ and $\pmb\gamma$, respectively.
The operator $\partial_{\epsilon_1\cdot\epsilon_n}$ transmutes $P_1$ to
\bea
\W P_1&=&-2(k_n\cdot k_1)\,{\cal A}_{\rm YMS}(1,n;\pmb{g}_{n+m}\setminus\{1,n\}|\vec{\pmb{\sigma}}_{n+m})\nn
& &-2\sum_{\vec{\pmb\a}}\,(k_n\cdot F_{\pmb\a}\cdot k_1)\,{\cal A}_{\rm YMS}(1,\vec{\pmb\a},n;
\pmb{g}_{n+m}\setminus\{\{1,n\}\cup\a\}|\vec{\pmb{\sigma}}_{n+m})\,,
\eea
which is proportional to the resulted object of replacing $\epsilon_1$ and $\epsilon_n$ by $k_1$ and $k_n$ in
${\cal A}_{\rm YM}(\vec{\pmb{\sigma}}_{n+m})$ \cite{Hu:2023lso}, thus vanishes due to the gauge invariance.

Then we need to deal with ${\cal T}[1,\cdots,n]P_2$. Based on the assumption that each $\pmb\gamma$ does not include $1$ or $n$,
and the definition of the operator ${\cal T}[1,\cdots,n]$, it is straightforward to find that the survived terms in $P_2$
which will not be annihilated by ${\cal T}[1,\cdots,n]$, are those $\pmb\gamma$ do not include any element in $\{1,\cdots,n\}$.
Consequently, the summation $\sum_{\vec{\pmb\gamma}}$ for ordered sets $\vec{\pmb\gamma}$ is reduced to $\sum_{\vec{\pmb\gamma}'}$,
where $\pmb\gamma'$ are sets of external legs satisfy $\{1,\cdots,n\}\cap\pmb\gamma'=\emptyset$.

To continue, we substitute
\bea
{\cal A}_{\rm YMS}(\vec{\pmb{\gamma}}';\pmb{g}_{n+m}\setminus\pmb\gamma'|\vec{\pmb{\sigma}}_{n+m})
={\cal T}[\vec{\pmb{\gamma}}']\,{\cal A}_{\rm YM}(\vec{\pmb{\sigma}}_{n+m})
\eea
into \eref{diff-P2} to get
\bea
{\cal T}[1,\cdots,n]\,P_2&=&{\cal T}[1,\cdots,n]\,\sum_{\vec{\pmb\gamma}'}\,
{\rm tr}(F_{\vec{\pmb\gamma}'})\,\Big({\cal T}[\vec{\pmb{\gamma}}']\,{\cal A}_{\rm YM}(\vec{\pmb{\sigma}}_{n+m})\Big)\nn
&=&\sum_{\vec{\pmb\gamma}'}\,{\rm tr}(F_{\vec{\pmb\gamma}'})\,
\Big({\cal T}[1,\cdots,n]\,{\cal T}[\vec{\pmb{\gamma}}']\,{\cal A}_{\rm YM}(\vec{\pmb{\sigma}}_{n+m})\Big)\,,
~~\label{TP2}
\eea
where the second equality uses the observation that ${\cal T}[1,\cdots,n]$ annihilates
${\rm tr}(F_{\vec{\pmb\gamma}'})$, due to $\{1,\cdots,n\}\cap\pmb\gamma'=\emptyset$.
By definition, two combinatorial operators ${\cal T}[1,\cdots,n]$ and ${\cal T}[\vec{\pmb{\gamma}}']$
transmutes ${\cal A}_{\rm YM}(\vec{\pmb{\sigma}}_{n+m})$ to double-trace YMS amplitude as
\bea
{\cal A}_{\rm YMS}(1,\cdots,n::\vec{\pmb\gamma}';\pmb{g}_{m}\setminus\pmb\gamma'|\vec{\pmb\sigma}_{n+m})
&=&{\cal T}[1,\cdots,n]\,{\cal T}[\vec{\pmb{\gamma}}']\,{\cal A}_{\rm YM}(\vec{\pmb{\sigma}}_{n+m})\,,
\eea
where two ordered sets associated with two traces are separated by the symbol $::$.
Using the expansion of double-trace YMS amplitudes provided in \cite{Du:2017gnh,Du:2024dwm}, and relabel $\vec{\pmb\gamma}'$ as $\vec{\pmb\a}_1$,
we arrive at the expansion formula in
\eref{n+mp-amp}.

According to the above discussion, one can also rewrite \eref{n+mp-amp} as
\bea
{\cal A}_{\rm YMS^{+2}}(1,\cdots,n;\pmb{g}_m|\vec{\pmb{\sigma}}_{n+m})
&=&\sum_{\vec{\pmb\gamma}'}\,{\rm tr}(F_{\vec{\pmb\gamma}'})\,
{\cal A}_{\rm YMS}(1,\cdots,n::\vec{\pmb\gamma}';\pmb{g}_{m}\setminus\pmb\gamma'|\vec{\pmb\sigma}_{n+m})\,.~~\label{n+mp-reform}
\eea
This new formula manifests the permutation invariance for external gluons, thus the permutation symmetries of \eref{n+3p-amp1}, \eref{n+3p-compact}
and \eref{n+mp-amp} are proved.

\section{Towards YM$^{+2h}$ amplitudes}
\label{sec-YM+2h}

This section extends the approach to construct YM$^{+2h}$ amplitudes with mass dimension exceeding ${\cal D}+2$.
Section \ref{subsec-YM+4-4p} and \ref{subsec-YM+4-np} focus on the special case with $h=2$, the resulted amplitudes correspond
to the Lagrangian
\bea
{\cal L}_{\rm YM^{+4}}=-{1\over4}\,F^{a}_{\mu\nu}F^{a\mu\nu}-{g\over 3\Lambda^2}\,f^{abc}F^{a~\nu}_{~\mu}F^{b~\rho}_{~\nu}F^{c~\mu}_{~\rho}
-{g^2\over4\Lambda^4}\,f^{abe}f^{ecd}\,F^a_{\mu\nu}F^b_{\rho\omega}F^{c\mu\nu}F^{d\rho\omega}\,,~~\label{Lag-+4}
\eea
which involves higher-derivative operators $F^3\equiv f^{abc}F^{a~\nu}_{~\mu}F^{b~\rho}_{~\nu}F^{c~\mu}_{~\rho}$ and
$F^4\equiv f^{abe}f^{ecd}\,F^a_{\mu\nu}F^b_{\rho\omega}F^{c\mu\nu}F^{d\rho\omega}$.  Our method naturally combines contributions from both of these operators. In Section \ref{subsec-h>3}, we put forward a conjecture for the general expansion formula that generates YM$^{\rm +2h}$ amplitudes from YM ones and verify the consistency of factorizations.

\subsection{$4$-point YM$^{+4}$ amplitude}
\label{subsec-YM+4-4p}

As discussed in section \ref{subsec-softbehavior} and \ref{subsec-YM+2-4p},
when considering the soft behavior at the $\tau^0$ order of $k_4\to \tau k_4$, the undetectable terms in
${\cal A}_{\rm YM^{+2}}(\vec{\pmb\sigma}_4)$ correspond to Feynman diagrams in which the leg $4$ is connected to a ${\rm YM}^{+2}$ vertex.
Using the full amplitude in \eref{4p-amp}, and the detectable part in \eref{4p-part} (with $\a=\b=1$), we see the undetectable part is
\bea
{\cal U}^4_{\rm YM^{+2}}(\vec{\pmb{\sigma}}_4)&=&{\cal A}_{\rm YM^{+2}}(\vec{\pmb{\sigma}}_4)-{\cal P}^4_{\rm YM^{+2}}(\vec{\pmb{\sigma}}_4)\nn
&=&\sum_{i=1}^3\,{\rm tr}(f_4\cdot f_i){\cal A}_{\rm YMS}(i,4;\{1,2,3\}\setminus i|\vec{\pmb{\sigma}}_4)\,,\label{eq:5.1-1}
\eea
which can be generated from the YM amplitude ${\cal A}_{\rm YM}(\pmb\sigma_4)$ as
\bea
{\cal U}^4_{\rm YM^{+2}}(\vec{\pmb{\sigma}}_4)&=&\Big(\sum_{i=1}^3\,{\rm tr}(f_4\cdot f_i)\,\partial_{\epsilon_4\cdot\epsilon_i}\Big)\,
{\cal A}_{\rm YM}(\vec{\pmb\sigma}_4)\,.
\eea
Thus, applying the operator $\left({\rm tr}(f_4\cdot f_i)\partial_{\epsilon_4\cdot\epsilon_i}\right)$ to ${\cal A}_{\rm YM}(\vec{\pmb\sigma}_4)$ creates a term where two external legs, $4$ and $i$, are connected to a ${\rm YM}^{+2}$ vertex, while the remaining two legs are connected to an ordinary YM vertex. Put differently, the operator $\left({\rm tr}(f_4\cdot f_i)\partial_{\epsilon_4\cdot\epsilon_i}\right)$ transform an ordinary YM vertex into a YM$^{+2}$ vertex (In the sense that each 4-point vertex has been decomposed into two trivalent vertices and a popagator). Next, let us consider
\bea
T_{(12)(34)}=\Big({\rm tr}(f_4\cdot f_3)\,\partial_{\epsilon_4\cdot\epsilon_3}\Big)\,
\Big({\rm tr}(f_2\cdot f_1)\,\partial_{\epsilon_2\cdot\epsilon_1}\Big)\,
{\cal A}_{\rm YM}(\vec{\pmb\sigma}_4)\,.~~\label{2operator}
\eea
Motivated by \eqref{eq:5.1-1}, $T_{(12)(34)}$ can be seen as a term featuring two ${\rm YM}^{+2}$ vertices: legs $4$ and $3$ link to one vertex, while legs $2$ and $1$ connect to the other. This statement is consist with the observation that $T_{(12)(34)}$ has only one pole $1/s_{12}$,
and factorizes to correct $3$-point ${\rm YM}^{+2}$ amplitudes near $s_{12}=0$. Thus, one can use $T_{(ab)(cd)}$ to construct
$4$-point ${\rm YM}^{+4}$ amplitudes with two ${\rm YM}^{+2}$ vertices, up to an overall constant $\lambda$,
\bea
{\cal A}_{\rm YM^{+4}}(\vec{\pmb\sigma}_4)&=&\lambda\,\sum_{\{1,b\}\cap\{c,d\}=\emptyset}\,T_{(1b)(cd)}\nn
&=&\lambda\,\sum_{\{1,b\}\cap\{c,d\}=\emptyset}\,\Big({\rm tr}(f_1\cdot f_b)\,\partial_{\epsilon_1\cdot\epsilon_b}\Big)\,
\Big({\rm tr}(f_c\cdot f_d)\,\partial_{\epsilon_c\cdot\epsilon_d}\Big)\,
{\cal A}_{\rm YM}(\vec{\pmb\sigma}_4)\,,~~\label{4p-YM+4-1}
\eea
where the summation for all $\{1,b\}\cap\{c,d\}=\emptyset$ ensures the invariance under the permutation of external legs.

Using the formula \eref{reform-4p-YMS+2}, it is direct to recognize that
\bea
T_{(ab)(cd)}={\rm tr}(f_b\cdot f_a)\,{\cal A}_{\rm YMS^{+2}}(a,b;\{c,d\}|\vec{\pmb\sigma}_4)
={\rm tr}(f_d\cdot f_c)\,{\cal A}_{\rm YMS^{+2}}(c,d;\{a,b\}|\vec{\pmb\sigma}_4)\,,
\eea
thus, for latter convenience, we reformulate \eref{4p-YM+4-1} as
\bea
{\cal A}_{\rm YM^{+4}}(\vec{\pmb\sigma}_4)&=&\sum_{\vec{\pmb\rho},|\pmb\rho|=2}\,{\rm tr}(F_{\vec{\pmb\rho}})\,
{\cal A}_{\rm YMS^{+2}}(\vec{\pmb\rho};\{g\}_4\setminus\pmb\rho|\vec{\pmb\sigma}_4)\,,~~\label{4p-YM+4}
\eea
where we have chosen $\lambda=2$ for simplicity. Since each ${\rm YMS}^{+2}$ amplitude contains at least two gluons and two scalars, the summation is restricted to those ordered sets $\pmb\rho$ contains precise two elements. Direct evaluation shows that each $4$-point ${\rm YMS}^{+4}$ amplitude in \eref{4p-YM+4} contains polynomials without any pole, such terms correspond to $4$-point interaction. This $4$-point interaction is recognized as the $F^4$ operator in the Lagrangian \eref{Lag-+4}. Our result in \eref{4p-YM+4} coincides with the expansion found in \cite{Bonnefoy:2023imz}, which is based on employing the classical equation of motion.

\subsection{General YM$^{+4}$ amplitudes}
\label{subsec-YM+4-np}

Start from \eref{4p-YM+4}, one can use the technique similar as in previous sections to construct the general ${\rm YM}^{+4}$ amplitudes. Let us take the $5$-point case as the example. The subleading soft theorem for YM amplitudes yields
\bea
{\cal A}^{(1)_5}_{\rm YM^{+4}}(\vec{\pmb\sigma}_5)&=&S^{(1)_5}_g\,{\cal A}_{\rm YM^{+4}}(\vec{\pmb\sigma}_5\setminus5)\nn
&=&S^{(1)_5}_g\,\Big(\sum_{\vec{\pmb\rho}}\,{\rm tr}(F_{\vec{\pmb\rho}})\,
{\cal A}_{\rm YMS^{+2}}(\vec{\pmb\rho};\{g\}_5\setminus\{\pmb\rho\cup5\}|\vec{\pmb\sigma}_5\setminus5)\Big)\nn
&=&\sum_{\substack{\vec{\pmb\rho},|{\pmb\rho}|=2\\5\not\in\pmb\rho}}\,{\rm tr}(F_{\vec{\pmb\rho}})\,
{\cal A}^{(1)_5}_{\rm YMS^{+2}}(\vec{\pmb\rho};\{g\}_5\setminus\pmb\rho|\vec{\pmb\sigma}_5)\nn
& &+\tau\,\sum_{\substack{\vec{\pmb\rho},|{\pmb\rho}|=3\\5\in\pmb\rho}}\,{\rm tr}(F_{\vec{\pmb\rho}})\,
{\cal A}^{(0)_5}_{\rm YMS^{+2}}(\vec{\pmb\rho};\{g\}_5\setminus\pmb\rho|\vec{\pmb\sigma}_5)\,,
\eea
which indicates
\bea
{\cal P}^5_{\rm YM^{+4}}(\vec{\pmb\sigma}_5)
&=&\sum_{\substack{\vec{\pmb\rho},|{\pmb\rho}|=2\\5\not\in\pmb\rho}}\,{\rm tr}(F_{\vec{\pmb\rho}})\,
{\cal A}_{\rm YMS^{+2}}(\vec{\pmb\rho};\{g\}_5\setminus\pmb\rho|\vec{\pmb\sigma}_5)\nn
& &+\sum_{\substack{\vec{\pmb\rho},|{\pmb\rho}|=3\\5\in\pmb\rho}}\,{\rm tr}(F_{\vec{\pmb\rho}})\,
{\cal A}_{\rm YMS^{+2}}(\vec{\pmb\rho};\{g\}_5\setminus\pmb\rho|\vec{\pmb\sigma}_5)\,.~~\label{5p-part}
\eea
Imposing the cyclic invariance, we get
\bea
{\cal A}_{\rm YM^{+4}}(\vec{\pmb\sigma}_5)
&=&\sum_{\vec{\pmb\rho},2\leq|\pmb\rho|\leq3}\,{\rm tr}(F_{\vec{\pmb\rho}})\,
{\cal A}_{\rm YMS^{+2}}(\vec{\pmb\rho};\{g\}_5\setminus\pmb\rho|\vec{\pmb\sigma}_5)\,.~~\label{5p-YM+4}
\eea
Comparing \eref{5p-YM+4} with \eref{5p-part}, we see that the undetectable part is
\bea
{\cal U}^5_{\rm YM^{+4}}(\vec{\pmb\sigma}_5)
&=&{\cal A}_{\rm YM^{+4}}(\vec{\pmb\sigma}_5)-{\cal P}^5_{\rm YM^{+4}}(\vec{\pmb\sigma}_5)\nn
&=&\sum_{\substack{\vec{\pmb\rho},|{\pmb\rho}|=2\\5\in\pmb\rho}}\,{\rm tr}(F_{\vec{\pmb\rho}})\,
{\cal A}_{\rm YMS^{+2}}(\vec{\pmb\rho};\{g\}_5\setminus\pmb\rho|\vec{\pmb\sigma}_5)\nn
& &+\sum_{\substack{\vec{\pmb\rho},|{\pmb\rho}|=3\\5\not\in\pmb\rho}}\,{\rm tr}(F_{\vec{\pmb\rho}})\,
{\cal A}_{\rm YMS^{+2}}(\vec{\pmb\rho};\{g\}_5\setminus\pmb\rho|\vec{\pmb\sigma}_5)\,.~~\label{5p-un}
\eea
The vanishing of these terms at the $\tau^0$ order can be verified as follows.
For the $|{\pmb\rho}|=2,5\in\pmb\rho$ part, the leading order of
${\rm tr}(F_{\vec{\pmb\rho}})$ is the $\tau^1$ order, and the soft behaviour of
${\cal A}_{\rm YMS^{+2}}(\vec{\pmb\rho};\{g\}_5\setminus\pmb\rho|\vec{\pmb\sigma}_5)$
vanishes at the $\tau^{-1}$ order since each $\pmb\rho$ includes only two scalars.
For the $|{\pmb\rho}|=3,5\not\in\pmb\rho$ part, the leading order of
${\rm tr}(F_{\vec{\pmb\rho}})$ is the $\tau^0$ order. Each set
$\{g\}_5\setminus\pmb\rho$ includes $2$ gluons, thus the soft behaviour of
${\cal A}_{\rm YMS^{+2}}(\vec{\pmb\rho};\{g\}_5\setminus\pmb\rho|\vec{\pmb\sigma}_5)$
vanishes at both $\tau^{-1}$ and $\tau^0$ orders, due to the formula of $(n+2)$-point ${\rm YMS}^{+2}$ amplitude in \eref{n+2p-amp}.

Repeating the above procedure, the general expansion formula is found to be
\bea
{\cal A}_{\rm YM^{+4}}(\vec{\pmb\sigma}_n)
&=&\sum_{\vec{\pmb\rho},2\leq|\pmb\rho|\leq n-2}\,{\rm tr}(F_{\vec{\pmb\rho}})\,
{\cal A}_{\rm YMS^{+2}}(\vec{\pmb\rho};\{g\}_n\setminus\pmb\rho|\vec{\pmb\sigma}_n)\,.~~\label{np-YM+4}
\eea
In \eref{5p-YM+4} and \eref{np-YM+4}, the summations are for all inequivalent $\vec{\pmb\rho}$,
under the constraint that each ${\rm YMS}^{+2}$ includes at least two external gluons and two external scalars. This condition
is introduced to emphasize the effect ordered sets $\vec{\pmb\rho}$, however, it is also reasonable to formally extend the region to all
$\pmb\rho\subset \pmb{g}_n$, since contributions from ineffective ones automatically vanish.
Using the formula in \eref{n+mp-reform}, we can also rewrite \eref{np-YM+4}
as
\bea
{\cal A}_{\rm YM^{+4}}(\vec{\pmb\sigma}_n)
&=&\sum_{\substack{\vec{\pmb\rho}_1,\vec{\pmb\rho}_2\\ \pmb\rho_1\cap\pmb\rho_2=\emptyset
}}\,{\rm tr}(F_{\vec{\pmb\rho}_1})\,{\rm tr}(F_{\vec{\pmb\rho}_2})\,
{\cal A}_{\rm YMS}(\vec{\pmb\rho}_1::\vec{\pmb\rho}_2;\{g\}_n\setminus\{\pmb\rho_1\cup\pmb\rho_2\}|\vec{\pmb\sigma}_n)\,,~~\label{np-YM+4-reform}
\eea
which expands ${\rm YM}^{+4}$ amplitudes to double-trace YM ones. Here an over counting from exchanging $\vec{\pmb\rho}_1$ and $\vec{\pmb\rho}_2$
occurs, motivates us to introduce a normalization factor $1/2$.
The above expansions of $n$-point amplitude in \eref{np-YM+4}
and \eref{np-YM+4-reform} look different from that found in \cite{Bonnefoy:2023imz}. In next subsection,
we will argue \eref{np-YM+4}
and \eref{np-YM+4-reform} satisfy consistent  factorizations.

\subsection{Conjecture of $h\geq3$ cases}
\label{subsec-h>3}

The expansion of ${\rm YM}^{+2}$ amplitudes, as presented in \eref{np-amp}, can be expressed as:
\bea
{\cal A}_{\rm YM^{+2}}(\vec{\pmb\sigma}_n)=\sum_{\vec{\pmb\rho},\pmb\rho\subset\pmb{g}_n}\,
{\rm tr}(F_{\vec{\pmb\rho}})\,\Big({\cal T}[\vec{\pmb\rho}]\,
{\cal A}_{\rm YM}(\vec{\pmb\sigma}_n)\Big)\,,
\eea
due to the transmutation relation \eref{trans-operator}, which converts YM amplitudes to YMS ones. Similarly, the expansion of ${\rm YM}^{+4}$ amplitude \eref{np-YM+4}, can be expressed as:
\bea
{\cal A}_{\rm YM^{+4}}(\vec{\pmb\sigma}_n)={1\over 2}\,\sum_{\vec{\pmb\rho},\pmb\rho\subset\pmb{g}_n}\,
{\rm tr}(F_{\vec{\pmb\rho}})\,\Big({\cal T}[\vec{\pmb\rho}]\,
{\cal A}_{\rm YM^{+2}}(\vec{\pmb\sigma}_n)\Big)\,,~~\label{+4}
\eea
since the transmutation relation in \eref{trans}. Here, we have introduced the normalization factor of $1/2$ in \eref{+4}, aiming to offset the overcounting issue discussed around \eref{np-YM+4-reform}. The formulas for both ${\rm YM}^{+2}$ and ${\rm YM}^{+4}$ cases exhibit strong similarity, suggesting a natural conjecture for the general ${\rm YM}^{+2h}$ amplitudes.
\bea
{\cal A}_{\rm YM^{+2h}}(\vec{\pmb\sigma}_n)={1\over h}\,\sum_{\vec{\pmb\rho},\pmb\rho\subset\pmb{g}_n}\,
{\rm tr}(F_{\vec{\pmb\rho}})\,\Big({\cal T}[\vec{\pmb\rho}]\,{\cal A}_{\rm YM^{+2(h-1)}}(\vec{\pmb\sigma}_n)\Big)\,.~~\label{YM+2h}
\eea
The normalization factor $1/h$ is once again introduced to prevent overcounting, as will be examined shortly. The general formula \eref{YM+2h} above possesses the correct mass dimension and manifests gauge and permutation invariance. This formula reveals a recursive pattern, enabling the construction of ${\rm YM}^{+2h}$ amplitudes for any value of $h$ from the YM ones, utilizing the differential operators ${\cal T}[\vec{\pmb\rho}]$ defined in \eref{defin-oper}. The summation is formally over all inequivalent ordered sets $\vec{\pmb\rho}$ with $\pmb\rho\subset\pmb{g}_n$, although the effective part constitutes a subset, as discussed below \eref{np-YM+4}.

This conjecture demonstrates two very favorable properties. The first one concerns the correct soft behaviors at leading and subleading orders. For any value of $h$, if we assume that the lowest-point amplitudes take the form described in \eref{YM+2h}, then the validity of \eref{YM+2h} for general YM$^{+2h}$ amplitudes naturally follows from the same recursive technique used to derive \eref{np-amp} and \eref{np-YM+4} based on the soft theorem.

Another property is the consistent factorizations. In section \ref{sec-YM+2}, we derived the expansion of YM$^{+2}$ amplitudes and proved the equivalence between the resulting expansion \eref{np-amp} and the expansion identified in \cite{Bonnefoy:2023imz} through covariant CK duality. Assuming the validity of the expansion \eref{np-amp} as the correct YM$^{+2}$ amplitudes with accurate factorizations, we can then establish that such accurate factorizations are inherited by the expansion formula \eref{YM+2h} for general YM$^{+2h}$ amplitudes.

To show this, let's begin by considering another expansion formula for YM$^{+2h}$ amplitudes. We define
\bea
{\cal B}(\vec{\pmb\rho};\pmb{g}_n\setminus\pmb\rho|\vec{\pmb\sigma}_n)
={\cal T}[\vec{\pmb\rho}]\,{\cal A}_{\rm YM^{+2h}}(\vec{\pmb\sigma}_n)\,.~~\label{defin-YMS+2h}
\eea
In \cite{Feng:2019tvb}, it was demonstrated that the transmutation, as described in \eref{defin-YMS+2h}, along with the requirement of gauge invariance, leads to the following expansion:
\bea
{\cal A}_{\rm YM^{+2h}}(\vec{\pmb\sigma}_n)=\sum_{\vec{\pmb\alpha},\pmb\a\subset\pmb{g}_n\setminus\{1,n\}}\,\big(\epsilon_n\cdot F_{\vec{\pmb\alpha}}\cdot\epsilon_1\big)\,
{\cal B}(1,\vec{\pmb\a},n;\pmb{g}_n\setminus\{\pmb\a\cup\{1,n\}\}|\vec{\pmb\sigma}_n)\,.~~\label{expan-to-YMS}
\eea
The derivation in \cite{Feng:2019tvb} is independent of the interpretation of ${\cal B}(1,\vec{\pmb\a},n;\pmb{g}_n\setminus\{\pmb\a\cup\{1,n\}\}|\vec{\pmb\sigma}_n)$ on the right-hand side of \eref{expan-to-YMS}\footnote{The operator ${\cal T}[\vec{\pmb\rho}]$ eliminates polarizations carried by gluons in the set $\pmb\rho$, while introducing an additional ordering denoted by $\vec{\pmb\rho}$. Hence, it's reasonable to interpret ${\cal T}[\vec{\pmb\rho}]{\cal A}_{\rm YM^{+2h}}(\vec{\pmb\sigma}_n)$ as the amplitude of gluons coupled to BAS scalars with mass dimension ${\cal D}'+2h$, namely, ${\cal A}_{\rm YMS^{+2h}}(\vec{\pmb\rho};\pmb{g}_n\setminus\pmb\rho|\vec{\pmb\sigma}_n)$. This conjectured interpretation is irrelevant to our discussion.}. This implies that the following expansion always holds:
\bea
{\cal A}_{\rm YM^{+2h}}(\vec{\pmb\sigma}_n)=\sum_{\vec{\pmb\alpha},\pmb\a\subset\pmb{g}_n\setminus\{1,n\}}\,
\big(\epsilon_n\cdot F_{\vec{\pmb\alpha}}\cdot\epsilon_1\big)\,\Big({\cal T}[1,\vec{\pmb\a},n]\,
{\cal A}_{\rm YM^{+2h}}(\vec{\pmb\sigma}_n)\Big)\,.~~\label{expan-to-YMS-abstract}
\eea
Therefore, we can define two operators
\bea
{\cal G}^{+2}(\pmb{g}_n)=\sum_{\vec{\pmb\rho},\pmb\rho\subset\pmb{g}_n}\,
{\rm tr}(F_{\vec{\pmb\rho}})\,{\cal T}[\vec{\pmb\rho}]\,,~~\label{G2}
\eea
and
\bea
{\cal G}(\pmb{g}_n)=\sum_{\vec{\pmb\alpha},\pmb\a\subset\pmb{g}_n\setminus\{1,n\}}\,
\big(\epsilon_n\cdot F_{\vec{\pmb\alpha}}\cdot\epsilon_1\big)\,{\cal T}[1,\vec{\pmb\a},n]\,.~~\label{G}
\eea
The first operator ${\cal G}^{+2}(\pmb{g}_n)$ transmutes YM$^{+2(h-1)}$ amplitudes to YM$^{+2h}$ ones,
while the second operator ${\cal G}(\pmb{g}_n)$
turns YM$^{+2(h-1)}$ amplitudes to themselves, according to relations \eref{YM+2h} and \eref{expan-to-YMS-abstract}.

Now we turn back to the factorization. With the assumption of the validity of expansion \eref{np-amp}, we have the following factorizations
which can be regarded as the input,
\bea
{\cal A}_{\rm YM}(\vec{\pmb\sigma}_n)\Big|_{s_{\pmb\b}\to0}=\sum_{\rm Inter}\,{\cal A}_{\rm YM}(\vec{\pmb\b},I_L)\,{1\over s_{\pmb\b}}\,
{\cal A}_{\rm YM}(\bar{\vec{\pmb\b}},I_R)\,,~~\label{fac-YM}
\eea
and
\bea
{\cal A}_{\rm YM^{+2}}(\vec{\pmb\sigma}_n)\Big|_{s_{\pmb\b}\to0}&=&\sum_{\rm Inter}\,{\cal A}_{\rm YM^{+2}}(\vec{\pmb\b},I_L)\,{1\over s_{\pmb\b}}\,
{\cal A}_{\rm YM}(\bar{\vec{\pmb\b}},I_R)\nn
& &+\sum_{\rm Inter}\,{\cal A}_{\rm YM}(\vec{\pmb\b},I_L)\,{1\over s_{\pmb\b}}\,
{\cal A}_{\rm YM^{+2}}(\bar{\vec{\pmb\b}},I_R)\,,~~\label{fac-YM+2}
\eea
where $I_L$ and $I_R$ encode particles arise from cutting internal line, and the summation is among all intermediate states
with different possible helicities. Two ordered sets
$\vec{\pmb\b}$ and $\bar{\vec{\pmb\b}}$ satisfy $\{\vec{\pmb\b},\bar{\vec{\pmb\b}}\}=\vec{\pmb\sigma}_n$, for instance
$\vec{\pmb\b}=\{2,3,4\}$, $\bar{\vec{\pmb\b}}=\{5,6,1\}$, $\vec{\pmb\sigma}_6=\{1,2,3,4,5,6\}$. Substituting operators \eref{G}
and \eref{G2} into \eref{fac-YM+2} gives
\bea
{\cal G}^{+2}(\pmb{g}_n)\,{\cal A}_{\rm YM}(\vec{\pmb\sigma}_n)\Big|_{s_{\pmb\b}\to0}
&=&\sum_{\rm Inter}\,\Big({\cal G}^{+2}(\pmb\b,I_L)\,{\cal A}_{\rm YM}(\pmb\b,I_L)\Big)\,{1\over s_{\pmb\b}}\,
\Big({\cal G}(\bar{\pmb\b},I_R)\,{\cal A}_{\rm YM}(\bar{\vec{\pmb\b}},I_R)\Big)\nn
& &+\sum_{\rm Inter}\,\Big({\cal G}(\pmb\b,I_L)\,{\cal A}_{\rm YM}(\vec{\pmb\b},I_L)\Big)\,{1\over s_{\pmb\b}}\,
\Big({\cal G}^{+2}(\bar{\pmb\b},I_R)\,{\cal A}_{\rm YM}(\bar{\vec{\pmb\b}},I_R)\Big)\,,~~\label{fac-YM+2-operator}
\eea
where $\pmb\b$ and $\bar{\pmb\b}$ are unordered sets of gluons correspond to ordered ones $\vec{\pmb\b}$ and $\bar{\vec{\pmb\b}}$,
respectively. Comparing \eref{fac-YM+2-operator} with \eref{fac-YM}, we observe the following
factorization of operator ${\cal G}^{+2}(\pmb{g}_n)$\footnote{The above factorization is not required to be valid at the algebraic level. We only expect this relation to hold when acting on
physical amplitudes.},
\bea
{\cal G}^{+2}(\pmb{g}_n)\Big|_{s_{\pmb\b}\to 0}={\cal G}^{+2}(\pmb\b,I_L)\,\otimes\,
{\cal G}(\bar{\pmb\b},I_R)+{\cal G}(\pmb\b,I_L)\,\otimes\,
{\cal G}^{+2}(\bar{\pmb\b},I_R)\,.~~\label{fac-operator}
\eea
By substituting the factorizations of YM$^{+2}$ amplitude in \eref{fac-YM+2} and operator ${\cal G}^{+2}(\pmb{g}_n)$ in \eref{fac-operator},
we immediately achieve
\bea
{\cal A}_{\rm YM^{+4}}(\vec{\pmb\sigma}_n)\Big|_{s_{\pmb\b}\to0}
&=&{1\over2}\,{\cal G}^{+2}(\pmb{g}_n)\,{\cal A}_{\rm YM^{+2}}(\vec{\pmb\sigma}_n)\Big|_{s_{\pmb\b}\to0}\nn
&=&{1\over2}\,\sum_{\rm Inter}\,\Big({\cal G}^{+2}(\pmb\b,I_L)\,{\cal A}_{\rm YM^{+2}}(\vec{\pmb\b},I_L)\Big)\,{1\over s_{\pmb\b}}\,
\Big({\cal G}(\bar{\pmb\b},I_R)\,{\cal A}_{\rm YM}(\bar{\vec{\pmb\b}},I_R)\Big)\nn
& &+{1\over2}\,\sum_{\rm Inter}\,\Big({\cal G}(\pmb\b,I_L)\,{\cal A}_{\rm YM^{+2}}(\vec{\pmb\b},I_L)\Big)\,{1\over s_{\pmb\b}}\,
\Big({\cal G}^{+2}(\bar{\pmb\b},I_R)\,{\cal A}_{\rm YM}(\bar{\vec{\pmb\b}},I_R)\Big)\nn
& &+{1\over2}\,\sum_{\rm Inter}\,\Big({\cal G}^{+2}(\pmb\b,I_L)\,{\cal A}_{\rm YM}(\vec{\pmb\b},I_L)\Big)\,{1\over s_{\pmb\b}}\,
\Big({\cal G}(\bar{\pmb\b},I_R)\,{\cal A}_{\rm YM^{+2}}(\bar{\vec{\pmb\b}},I_R)\Big)\nn
& &+{1\over2}\,\sum_{\rm Inter}\,\Big({\cal G}(\pmb\b,I_L)\,{\cal A}_{\rm YM}(\vec{\pmb\b},I_L)\Big)\,{1\over s_{\pmb\b}}\,
\Big({\cal G}^{+2}(\bar{\pmb\b},I_R)\,{\cal A}_{\rm YM^{+2}}(\bar{\vec{\pmb\b}},I_R)\Big)\nn
&=&\sum_{\rm Inter}\,{\cal A}_{\rm YM^{+4}}(\vec{\pmb\b},I_L)\,{1\over s_{\pmb\b}}\,
{\cal A}_{\rm YM}(\bar{\vec{\pmb\b}},I_R)\nn
& &+\sum_{\rm Inter}\,{\cal A}_{\rm YM^{+2}}(\vec{\pmb\b},I_L)\,{1\over s_{\pmb\b}}\,
{\cal A}_{\rm YM^{+2}}(\bar{\vec{\pmb\b}},I_R)\nn
& &+\sum_{\rm Inter}\,{\cal A}_{\rm YM}(\vec{\pmb\b},I_L)\,{1\over s_{\pmb\b}}\,
{\cal A}_{\rm YM^{+4}}(\bar{\vec{\pmb\b}},I_R)\,,
\eea
which is the expected factorization of YM$^{+4}$ amplitude
near $s_{\pmb\b}= 0$.
Repeating the above process, it is straightforward to show the general factorization
\bea
{\cal A}_{\rm YM^{+2h}}(\vec{\pmb\sigma}_n)\Big|_{s_{\pmb\b}\to0}
=\sum_{\ell=0}^h\,\sum_{\rm Inter}\,{\cal A}_{\rm YM^{+2\ell}}(\vec{\pmb\b},I_L)\,{1\over s_{\pmb\b}}\,
{\cal A}_{\rm YM^{2(h-\ell)}}(\bar{\vec{\pmb\b}},I_R)\,.~~\label{fac-general}
\eea

By substituting \eref{YM+2h} into itself iteratively, we obtain
\bea
{\cal A}_{\rm YM^{+2h}}(\vec{\pmb\sigma}_n)&=&{1\over h!}\,\sum_{\substack{\vec{\pmb\rho}_1,\cdots,\vec{\pmb\rho}_h\\ \pmb\rho_a\cap
\pmb\rho_b=\emptyset}}\,
\Big(\prod_{j=1}^h\,{\rm tr}(F_{\vec{\pmb\rho}_j})\Big)\,\Big({\cal T}[\vec{\pmb\rho}_1]\cdots
{\cal T}[\vec{\pmb\rho}_h]\,{\cal A}_{\rm YM}(\vec{\pmb\sigma}_n)\Big)\nn
&=&{1\over h!}\,\sum_{\substack{\vec{\pmb\rho}_1,\cdots,\vec{\pmb\rho}_h\\ \pmb\rho_a\cap
\pmb\rho_b=\emptyset}}\,
\Big(\prod_{j=1}^h\,{\rm tr}(F_{\vec{\pmb\rho}_j})\Big)\,{\cal A}_{\rm YMS}(\vec{\pmb\rho}_1::\cdots::\vec{\pmb\rho}_h;
\pmb{g}_n\setminus\{\pmb\rho_1\cup\cdots\cup\pmb\rho_h\}|\vec{\pmb\sigma}_n)\,,~~\label{YM+2h-reform}
\eea
which is the generalization of expansion \eref{np-YM+4-reform}. When deriving \eref{np-YM+4-reform}, we utilized the expansion of YM amplitudes as detailed in \cite{Hu:2023lso}. This expansion suggests that the $P_1$ component defined in \eref{diff-P1} is annihilated by the operator $\partial_{\epsilon_1\cdot\epsilon_n}$. Consequently, we have the expansion \eref{n+mp-reform}, which subsequently leads to \eref{np-YM+4-reform}. The derivation for \eref{YM+2h-reform} follows a similar methodology, with the analogous expansion of YM$^{+2h}$ amplitudes in \eref{expan-to-YMS-abstract} assuming the role of the expansion of YM amplitudes in \cite{Hu:2023lso}. The first line on the right-hand side of \eref{YM+2h-reform} elucidates the process of generating general YM$^{+2h}$ amplitudes from YM ones, independent of the representation of YM amplitudes. The second line is crucial for expanding YM$^{+2h}$ amplitudes to BAS amplitudes, given that the expansion of multi-trace YMS amplitudes is already established \cite{Du:2017gnh,Du:2024dwm}. From the expansion \eref{YM+2h-reform}, it is evident that the normalization factor $1/h!$ mitigates the overcounting resulting from exchanging $\vec{\pmb\rho}_a$.

\begin{figure}
  \centering
  \includegraphics[width=12cm]{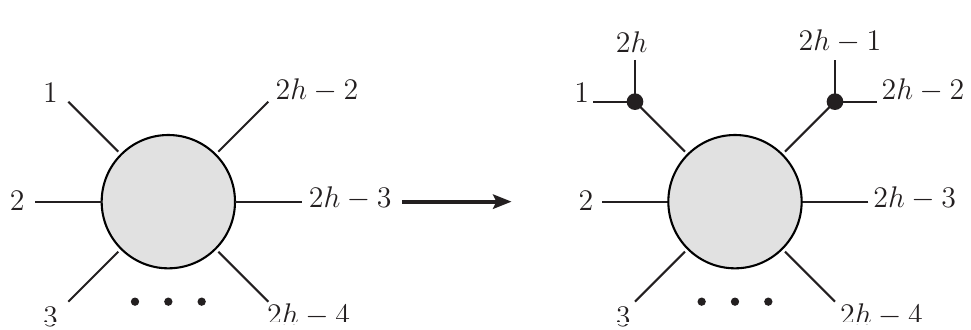} \\
  \caption{Turn $2(h-2)$-point YM$^{+2(h-2)}$ amplitude to $2h$-point YM$^2h$ amplitude, via two YM$^{+2}$ vertices.}\label{2h}
\end{figure}

Before concluding this subsection, we provide some remarks regarding the $2h$-point YM$^{+2h}$ amplitudes. Naively, the lowest-point YM$^{+2h}$ amplitude consists of $(h+2)$ external legs if all $n-2$ vertices in an $n$-point trivalent diagram are YM$^{+2}$ ones. Our formula above yields amplitudes with at least $2h$ external legs. This is because that YM$^{+2h}$ amplitudes in \eref{YM+2h} and \eref{YM+2h-reform} consistently involve $2h$-point interactions. As illustrated in Figure \ref{2h}, one can insert two additional external gluons attached to two trivalent YM$^{+2}$ vertices to generate a $2h$-point term with mass dimension ${\cal D}+2h$ from the $2(h-1)$-point YM$^{+2(h-1)}$ amplitude. Since the factor ${\rm tr}(f_1\cdot f_{2h}){\rm tr}(f_{2h-1}\cdot f_{2h-2})$ contains the term $(\epsilon_1\cdot\epsilon_{2h})(\epsilon_{2h-1}\cdot\epsilon_{2h-2})s_{1(2h)}s_{(2h-1)(2h-2)}$, which cancels propagators $1/s_{1(2h)}$ and $1/s_{(2h-1)(2h-2)}$, we observe that the $2h$-point YM$^{+2h}$ amplitude contains polynomials without any poles, if the associated $2(h-1)$-point YM$^{+2(h-1)}$ amplitude contains polynomials. Since $4$-point YM$^{+4}$ amplitudes contain $F^4$ vertices, the above argument recursively ensures the existence of $2h$-point interactions with any value of $h$. We have verified this observation by evaluating
examples, up to $h = 10$.
Thus, we observe an interesting phenomenon: while the Lagrangian includes an infinite number of higher-derivative interactions and is consequently very complex, the associated scattering amplitudes exhibit a remarkably compact and universal structure in \eref{YM+2h} and \eref{YM+2h-reform}.

%
%

\section{Summary}
\label{sec-summary}

In this paper, we extend the bottom-up approach developed in our recent works to effective field theories encompassing gluons and BAS scalars. By inverting the subleading soft theorem for YM amplitudes, we construct the tree ${\rm YM}^{+2}$ amplitudes featuring a single insertion of the ${\rm YM}^{+2}$ vertex. This construction benefits from the observation that the subleading soft theorem for YM amplitudes remains valid for ${\rm YM}^{+2}$ amplitudes. However, the direct construction only captures a subset of the complete amplitude, with the remainder identified through the consideration of appropriate symmetries. Subsequently, we proceed to construct the tree ${\rm YMS}^{+2}$ amplitudes by employing the inversion of the leading soft theorem for BAS amplitudes and the subleading soft theorem for YM amplitudes, guided by a similar rationale. Finally, we delve into the realm of tree ${\rm YM}^{+2h}$ amplitudes with higher values of $h$. We derive the complete result for $h=2$ and propose a conjecture for the general formula applicable to any value of $h$. We also discuss the consistent factorizations of the conjectured general formula. Importantly, all constructions in this paper are independent of the Lagrangian or Feynman rules, and the potential infinite tower of higher-derivative operators are automatically encompassed.

As a direct consequence of our method, all resulting amplitudes are expressed as universal expansions to YMS$^{+2h'}$ amplitudes with lower $h'$ or expansions to multiple-trace YMS amplitudes, maintaining manifest permutation and gauge invariance. Furthermore, in the expansion of ${\cal A}_{\rm YM}^{+2h}(\vec{\pmb\sigma}_n)$ or ${\cal A}_{\rm YMS}^{+2h}(1,\cdots,n;\pmb{g}_m|\vec{\pmb\sigma}_{n+m})$, the coefficients are independent of the orderings $\vec{\pmb\sigma}_n$ or $\vec{\pmb\sigma}_{n+m}$. As noted at the end of section \ref{subsec-YM+2-4p}, this feature is equivalent to the well-known double copy structure. Ultimately, these expansions naturally lead to the representations of these amplitudes as expansions to pure BAS ones, ensuring their adherence to the BCJ relations.

There are various aspects to further understand YM$^{+2h}$ amplitudes with $h\geq3$. Firstly, as discussed at the end of section \ref{subsec-YM+2-np}, although the expansions exhibit gauge and permutation invariance, they are highly redundant. Therefore, it is important to discover alternative formulas that eliminate these redundancies. Secondly, it is intriguing to determine the Lagrangian corresponding to YM$^{+2h}$ amplitudes for each $h\geq3$, as well as the correspondence between each term and Feynman diagrams. Thirdly, since the expansions identified in this paper display a manifest double-copy structure, another significant task is to search for the corresponding CHY formulas, serving as generalizations of the CHY formula for YM$^{+2}$ amplitudes found in \cite{He:2016iqi}. Finally, assuming we can discover the correct formulas, the next essential question is their uniqueness.

We conclude with a brief discussion on potential generalizations of our results and method. Firstly, following the double copy structure, it is natural to anticipate that the expansions found in \eref{np-amp}, \eref{n+mp-amp}, and \eref{np-YM+4} can be directly mapped to gravitational (GR) and Einstein-Yang-Mills (EYM) amplitudes with the same coefficients. The resulting GR and EYM amplitudes will naturally be interpreted as those with higher-derivative interactions, akin to Gauss-Bonnet gravity arising from the low-energy effective action of closed strings, among others. Secondly, since the soft limit can be taken for any massless particle, it is intriguing to investigate the applicability of our method in a broader scope, such as scalar effective field theories like the non-linear sigma model and beyond, as well as amplitudes of theories in which massless particles couple to fermions.

\section*{Acknowledgments}

KZ is supported by NSFC under Grant No. 11805163.


\bibliographystyle{JHEP}

\bibliography{reference}

\end{document}